\documentclass[aps,preprint,footinbib,superscriptaddress]{revtex4-2}
\usepackage[margin=1in]{geometry}
\usepackage[group-separator={,}]{siunitx}
\usepackage{dcolumn,amsmath,setspace,graphicx,comment,amssymb,bm,natbib,caption,subcaption}

\newcommand{\be}{\begin{equation}}
\newcommand{\ee}{\end{equation}}

\begin{document}
\title{Generation of momentum transport in weakly turbulent $\beta$-plane magnetohydrodynamics}
\author{R.\ A.\ Heinonen}
\affiliation{Dept.\ Physics and INFN, University of Rome ``Tor Vergata'', Via della Ricerca Scientifica 1, 00133 Rome, Italy}
\author{P.\ H.\ Diamond}
\affiliation{Dept.\ Physics, University of California San Diego, La Jolla, California 92093, USA}
\author{M.\ F.\ D.\ Katz} 
\affiliation{Dept.\ Physics and INFN, University of Rome ``Tor Vergata'', Via della Ricerca Scientifica 1, 00133 Rome, Italy}
\author{G.\ E.\ Ronimo}
\affiliation{Dept.\ Physics, University of California San Diego, La Jolla, California 92093, USA}
\date{\today}
\begin{abstract}Magnetohydrodynamic (MHD) turbulence on a $\beta$-plane with an in-plane mean field, a system which serves as a simple model for the solar tachocline, is investigated analytically and computationally. We first derive two useful analytic constraints: we express the mean turbulent cross-helicity in terms of the mean turbulent magnetic energy, and then show that (for weak turbulence) the time-averaged momentum transport in the system can be expressed in terms of the cross-helicity spectrum. We then complete a closure of the system using weak turbulence theory, appropriately extended to a system with multiple interacting eigenmodes. We use this closure to perturbatively solve for the spectra at lowest order in the Rossby parameter $\beta$ and thereby show that the momentum transport in the system is $O(\beta^2)$, thus quantifying the transition away from Alfv\'enized turbulence. Finally, we verify our theoretical results by performing direct numerical simulations of the system over a broad range of $\beta$. \end{abstract}
\maketitle
\section{Introduction}\label{sec:intro}
Since the early 1990s, much attention has been paid to the \emph{solar tachocline}, a thin, stratified layer which lies at the base of the convective zone (CZ) of the sun. The tachocline separates the core (which rigidly rotates on spheres) from the CZ (which differentially rotates on cylinders) and exhibits strong radial shear (see, for example, \cite{schou98}). It is widely believed \cite{charbonneau2020,miesch,hughes_book} that the tachocline is home to the so-called $\Omega$-effect and thus is of crucial importance to the solar dynamo: the shear drags poloidal magnetic field lines originating from the core, converting them to a strong toroidal field, which is stored in the tachocline. According to Parker's ``interface dynamo'' theory \cite{parker93}, the $\alpha$-effect, whereby small-scale, turbulent helical motions twist the toroidal field back into the poloidal direction, occurs just above the tachocline where the toroidal field is stored, at the bottom of the CZ, preventing the stored field from quenching the $\alpha$-effect \cite{cattaneo96}.

Considerable debate in the literature has surrounded the nature of turbulent momentum transport in the tachocline, which is of central importance to the question of why the tachocline does not tend to spread into the inner radiative zone due to the motion of meridional cells in the CZ. Spiegel and Zahn (1992) \cite{spiegel92} first proposed that the turbulence would act as an eddy viscosity which balances the inward radial transport of angular momentum on large scales. Gough and McIntyre (1998) \cite{gough98} instead argued that the tachocline is essentially 2D due to strong stratification (see \cite{garaud} for a more recent perspective), and so turbulence would act as a \emph{negative} viscosity and produce mean ordered jets through the mixing of potential vorticity (PV). In the presence of differential rotation, PV (as opposed to angular momentum) is the essential conserved quantity, and it is defined as ${\rm PV}=\zeta+f,$ where $\zeta$ is the vorticity and $f$ the Coriolis parameter. How such PV mixing would oppose the downward ``burrowing'' of the tachocline remains unclear. The presence of the toroidal magnetic field is also known to play a profound role in this story as well: the field gives rise to magnetohydrodynamic (MHD) instabilities \cite{dikpati1999} which can redistribute angular momentum \cite{cally2001} and quench the formation of jets \cite{tobias2007}.

It is thus fruitful to study turbulent momentum transport in the tachocline, and in particular how it is affected by differential rotation and the magnetic field. Because the lower tachocline is stably stratified, it can be modeled as a quasi-two-dimensional, differentially rotating magnetic fluid. We assume the presence of a relatively strong toroidal field, and consider an MHD model on a 2-D $\beta$-plane:
\begin{align}
\label{eq:mhd1}\partial_t \nabla^2 \psi& + \beta \partial_x \psi = \{\psi, \nabla^2 \psi \} - \{ A, \nabla^2 A \} + \nu \nabla^4 \psi + \tilde f \\
\label{eq:mhd2}\partial_t A& = \{ \psi, A\} + \eta \nabla^2 A + \tilde g.
\end{align}
These equations express the dynamics of the streamfunction $\psi$ (and the vorticity $\nabla^2 \psi$) and the magnetic potential $A$ of an incompressible fluid. $A$ is an active scalar transported by the flow, inducing momentum transport (or equivalently, vorticity transport) through the Maxwell stress term $\{ A, \nabla^2 A \} $. The other nonlinear term generating momentum transport is the Reynolds stress term $\{ A, \nabla^2 A \} $, which describes the self-convection of the flow. We have defined the Poisson bracket $\{ a, b\} = \partial_x a \partial_y b - \partial_y a \partial_x b$, which acts as a convective derivative. The equations can be straightforwardly derived from the full incompressible 3-D MHD equations with a Coriolis force term by imposing planar velocity and magnetic fields $\mathbf{v} = ( \partial_y \psi, -\partial_x \psi, 0)$ and $\mathbf{b} = ( \partial_y A, -\partial_x A, 0)$, respectively, followed by taking the curl of the momentum equation and ``uncurling'' the magnetic induction equation. $\beta$ is the so-called Rossby parameter, which captures the effect of the differential rotation, i.e., there is a planetary vorticity gradient yielding a latitude-dependent Coriolis force. Explicitly, the solar rotation rate is locally given by $2 \bm{\Omega} \simeq ( 0 , 0, f_0+ \beta y)$. $\nu$ is the fluid viscosity and $\eta$ is the resistivity, both nondimensional. At the moment we do not specify the forcing functions $\tilde f$ and $\tilde g$, which respectively model kinetic stirring (including thermal excitation) and injection of magnetic potential, say by pumping from above or below the tachocline. It is worth noting that we have normalized the magnetic field by $1/\sqrt{4 \pi \rho},$ where $\rho$ is the density, so that it has units of velocity (and the Alfv\'en speed is unity).

This is probably the simplest possible model for the tachocline. The shallow-water MHD (SMHD) equations \cite{gilman2000,schecter2001} comprise one example of a more complex model, which can be obtained by relaxing the restriction to two dimensions, instead allowing slow variation along the $\hat{z}$ direction and including gravitational effects. Despite its simplicity, the beta-plane model exhibits a variety of interesting phenomena, including turbulence generated by the interaction of multiple modes, a transport bifurcation, and the advection of an active scalar; it is arguably of interest in its own right as a toy model for studying these phenomena and their interplay.

While introduced as a model of the tachocline, we note a second application of the beta-plane MHD system is to magnetically confined fusion, where, after minor modifications (that is, restoring a finite ion gyromagnetic scale $\rho_s$ to the vorticity equation), we recover a version of the Charney-Hasegawa-Mima equation \cite{charney,hm77} coupling drift waves with magnetic fluctuations.  There, the analog of the Rossby number is the mean pressure gradient. The strong toroidal field approximately confines the plasma to motion in the perpendicular plane, and drift-wave turbulence generated by the pressure gradient instability interacts with magnetic turbulence. This interaction is of intense interest due to the fact resonant magnetic perturbations, a technique used to stabilize transport at the edge, can weaken or destroy the confinement-enhancing shear flow which is generated by the turbulence --- see, for example, \cite{kriete2020}.

There has been substantial previous research on the beta-plane MHD model. Introduced by Tobias \emph{et al.}\ (2007) \cite{tobias2007} and Diamond \emph{et al.}\ (2007) \cite{diamond2007}, it constitutes a major simplification of the physics of the tachocline, but nevertheless features rich dynamics. Tobias \emph{et al.}\ studied the effect of a weak mean magnetic field aligned with the shear, so that $ A \to \tilde A + b_0 y$, and showed numerically that, when the resistivity becomes sufficiently large (or the mean field becomes sufficiently weak), the system undergoes a transition from Alfv\'enic turbulence with a forward cascade to Rossby-like turbulence exhibiting an inverse cascade to large scales. The inverse cascade is accompanied by zonal flow formation. 

\begin{figure}
\centering
\includegraphics[width=0.7\textwidth]{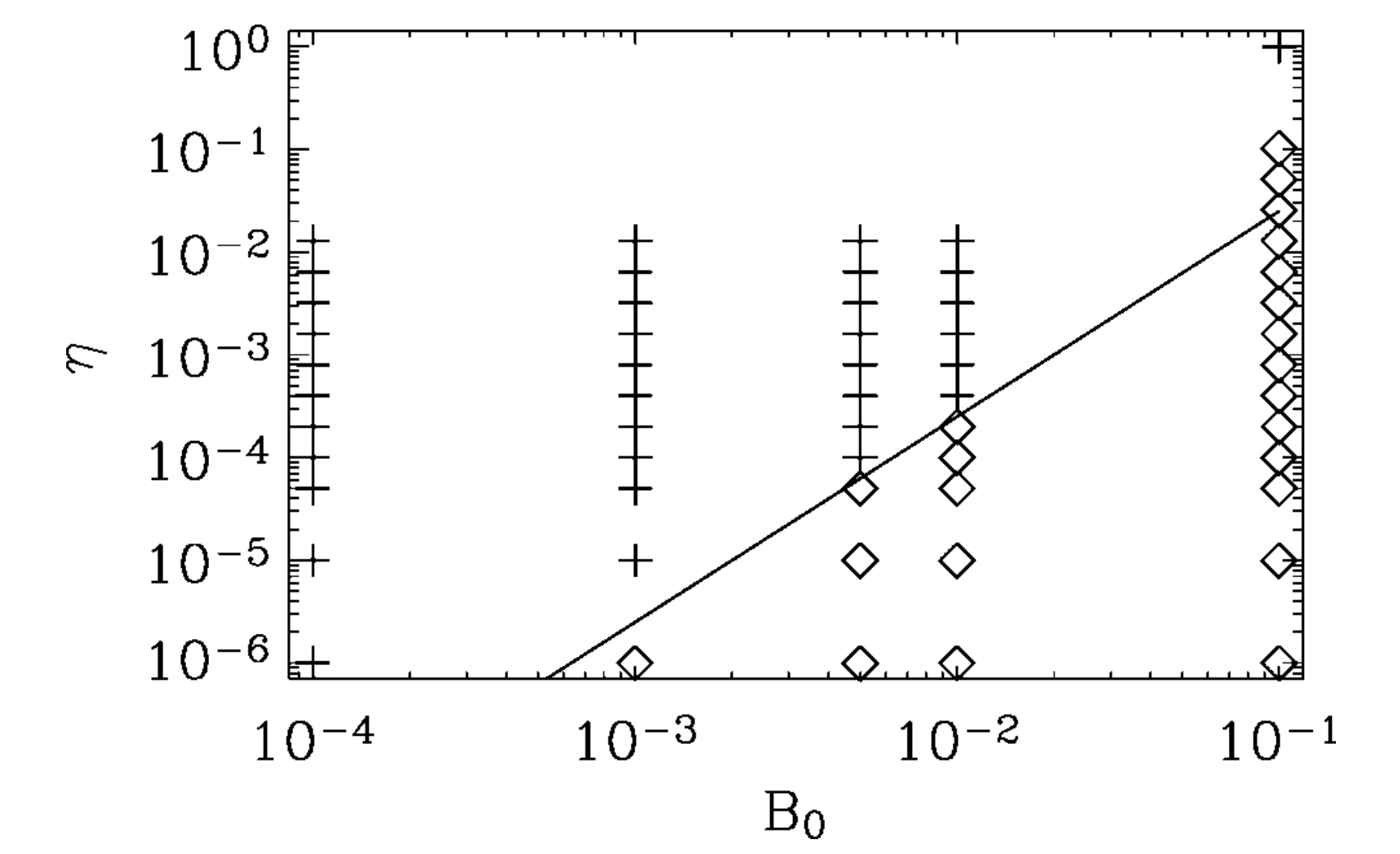}
\caption{In the $\beta$-plane MHD system, the line $b_0^2/\eta={\rm const.}$ separates the $(b_0, \eta)$ plane into two regions, one marked by an inverse cascade (plus signs) and one by a direct cascade (diamonds). Reproduced with permission from \cite{tobias2007}.}
\label{fig:etal}
\end{figure}

In particular, Tobias \emph{et al.}\ explored the dependence of zonal flow formation on the Zel'dovich parameter $Z={\rm Rm} v_{A,0}^2/\langle \tilde v^2 \rangle.$ Here, ${\rm Rm}$ is the magnetic Reynolds number, $v_{A,0}$ is the Alfv\'en speed in the externally prescribed magnetic field, and $\langle \tilde v^2 \rangle$ is the mean square velocity fluctuation intensity. Extensive previous work \cite{cattaneo91,gruzinov94,gruzinov96,fan2019} noted that for $Z<1,$ turbulent resistivity tracked kinematic estimates, while for $Z>1,$ the turbulent resistivity was quenched. A key point of $Z$-induced quenching is that only a weak $b_0$ field is needed at large ${\rm Rm}.$ The physics underlying quenching can be understood as vortex disruption \cite{mgh}. Given the previous history, it was natural to investigate $\beta$-plane MHD by a scan of $Z.$ Noting that $Z= \tau_c v_{A,0}^2/ \eta,$ where $\tau_c$ is the (fixed) forcing cell correlation time and $\eta$ the resistivity, Tobias \emph{et al.}\ scanned the parameter space of $b_0^2$ and $\eta,$ so as to examine jet formation. Results indicate that the $(b_0, \eta)$ plane divides into regions above and below a line defined by $b_0^2/ \eta = {\rm const.}$ (see Fig.~\ref{fig:etal}). Above the line, $b_0$ is strong, and jets are suppressed. Below the line, jets form. Alternatively, strong $b_0$ clearly inhibits jet formation via Alfv\'enization, while $\eta$ dissipates $b_0,$ and thus removes the inhibitor. The manifest $\eta$ dependence supports the relevance of the Zel'dovich parameter. The $b_0^2/ \eta = {\rm const.}$ scaling was also recovered analytically in \cite{parker}. 

\begin{figure}
\centering
\includegraphics[width=0.7\textwidth]{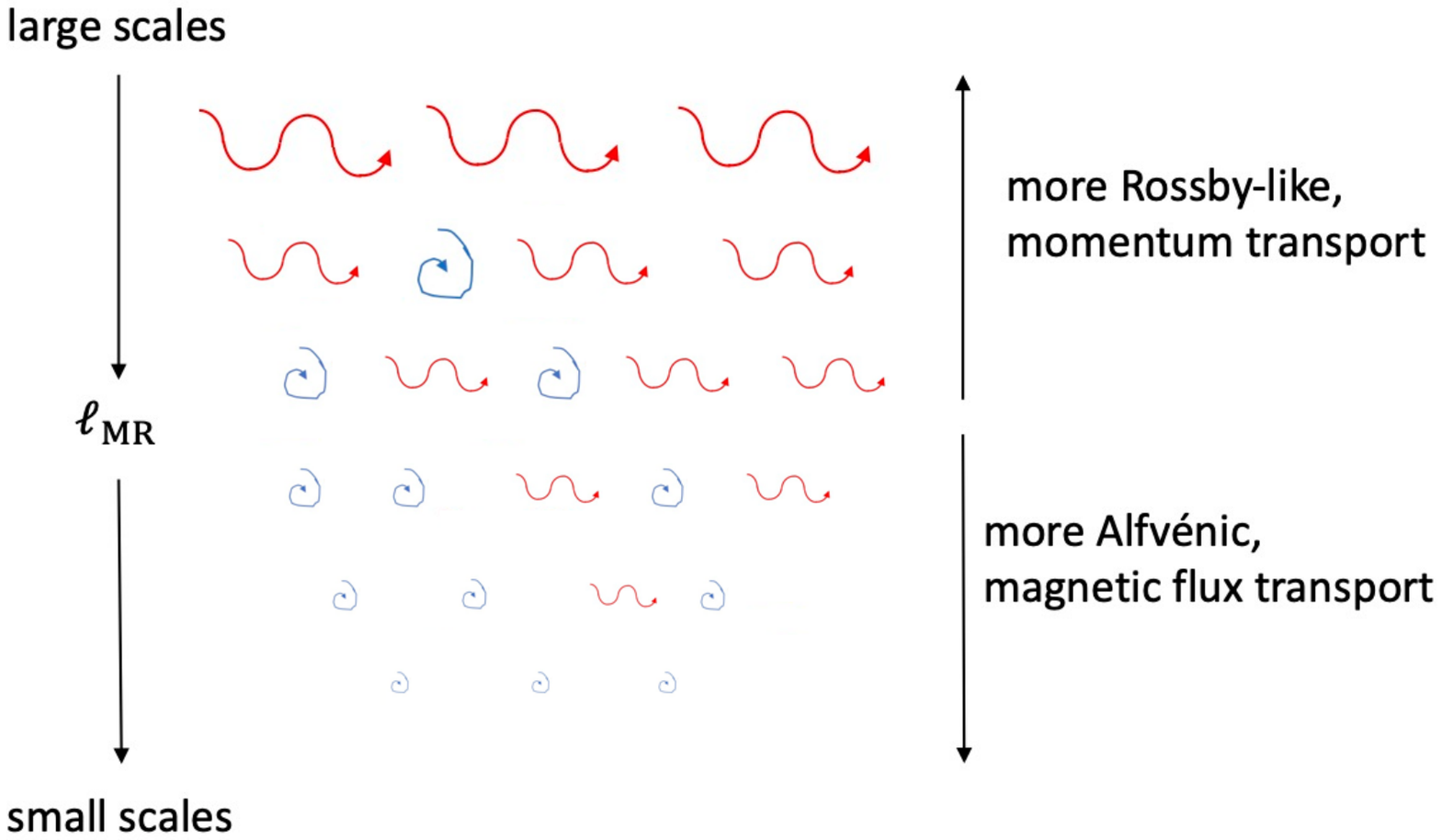}
\caption{Cartoon illustrating how the turbulence is separated by the magnetic Rhines scale. At larger scales, the turbulence is increasingly dominated by Rossby-like fluctuations (red squiggles) which transport momentum but do not transport magnetic field lines. At smaller scales, the turbulence is instead dominated by Alfv\'enic like waves (blue spirals) which do not transport momentum but carry magnetic flux.}
\label{fig:cartoon}
\end{figure}

Diamond \emph{et al.}\ (2007) gave an analytic closure calculation and derived the ``magnetic Rhines scale'' $\ell_{\rm MR} = \sqrt{b_0/\beta}$, defined by the crossover of the Rossby frequency $\omega_\beta = -\beta k_x/k^2$ and the Alfv\'en frequency $\omega_A = k_x b_0.$ They argued that $\ell_{\rm MR}$ is a critical lengthscale below which the turbulence is Alfv\'enized, so that the Maxwell-Reynolds stress
\begin{equation}
\langle \partial_x \tilde \psi \partial_y\tilde  \psi \rangle - \langle \partial_x \tilde A\partial_y \tilde A \rangle \simeq \int d^2\mathbf{k} \, \frac{k_x k_y}{k^2} (|\tilde{\mathbf{v}}_\mathbf{k}|^2 - |\tilde{\mathbf{b}}_\mathbf{k}|^2)
\end{equation}
tends to vanish to leading order and thus there is no clear large-scale momentum transport to set up a zonal flow (throughout this paper, $\langle f \rangle$ is understood to be a spatial average when $f$ is a real-space function and an ensemble average when $f$ is a $\mathbf{k}$-space function, unless explicitly stated otherwise). Conversely, at scales larger than the magnetic Rhines scale, the turbulence is Rossby-like --- see Fig.~\ref{fig:cartoon}. Thus, it is natural to suspect that the system will undergo transition from Alfv\'enic to Rossby turbulence when the magnetic Rhines scale becomes smaller than typical scales of the problem. 

Subsequent work \cite{chen2020} addressed the key question of the physics of jet quenching by MHD fluctuations. One candidate mechanism is the competition between Reynolds and Maxwell stresses, which tend to cancel in the Alfv\'enic limit. Another is decoherence of the Reynolds (and Maxwell) stresses by magnetic fields. Results indicate that both mechanisms are at work, with decoherence setting in at $b_0$ levels below that required for Alfv\'enization. A major motivation for the present work is to explore the effect of finite cross-helicity on the Reynolds vs.\ Maxwell competition. 

We note in passing that Guervilly and Hughes \cite{guervilly2015,guervilly2017} have shown in 3D that a small scale magnetic dynamo can inhibit large scale jet formation. In particular, they showed that when the magnetic Prandtl number exceeds the critical value for the onset of a small scale dynamo, the Reynolds stress decoheres and jet formation weakens. This is a related example of magnetic-field-induced inhibition of jet formation. 

However, omitted from previous studies is a reckoning with the fact that the Coriolis term explicitly breaks conservation of cross-helicity, that is the alignment of the magnetic field and velocity $H=\langle \tilde{\mathbf{v}} \cdot \tilde{\mathbf{b}} \rangle = - \langle\tilde  A \nabla^2 \tilde \psi \rangle$. Cross-helicity is globally conserved in pure MHD turbulence (up to dissipation), but when $\beta \ne 0$ we have
\begin{equation} \partial_t \langle \tilde A \nabla^2 \tilde \psi \rangle = -\beta \langle \tilde v_y \tilde A \rangle + {\rm dissipation}. \end{equation} The origin of the breaking of cross-helicity conservation is that the $\beta$ term, which (unlike a mean magnetic field term) cannot be removed by a simple change of variables, breaks symmetry in the $\hat{x}$ direction.

We can anticipate that a nonzero cross-helicity in this system may have important consequences for transport. Cross-helicity is closely related to the turbulent emf \cite{moffatt78}, so that one may naturally suspect that the cross-helicity induced by differential rotation in the tachocline has consequences for the solar dynamo (see \cite{yokoi} for more on the relationship between cross-helicity and the dynamo). However, it is well known that a dynamo cannot be supported in two-dimensions, so our primary subject of study will be momentum transport, which will generally couple to the cross-helicity in this system. In fact, we will see that the relationship between cross-helicity and momentum transport in this system is quite intimate.  

The link between cross helicity and transport may be expressed in terms of the spectrum $H_\mathbf{k} = \langle\tilde{ \mathbf{v}}_\mathbf{k} \cdot \tilde{ \mathbf{b}}_\mathbf{-k} \rangle.$ We then have
\begin{align}
\frac{k_x k_y}{2k^2}  \operatorname{Re} H_\mathbf{k} &=  \hat{z} \cdot \langle \tilde{\mathbf{v}}_\mathbf{k} \times \tilde{\mathbf{b}}_\mathbf{-k} \rangle \\
\frac{k_x}{k^2}  \operatorname{Im} H_\mathbf{k} &= \langle \tilde v_{y,\mathbf{k}} \tilde A_\mathbf{-k} \rangle.
\end{align}
Thus, the real part of the cross helicity determines the turbulent emf, and the imaginary part determines the turbulent flux of magnetic potential along the perpendicular axis. The turbulent emf is responsible for mean field evolution.

A useful ``dual'' quantity to the cross-helicity --- the alignment of the velocity field and magnetic field --- is the alignment of \emph{Els\"asser} populations. The Els\"asser basis consists of the variables $\mathbf{z}^\pm = \tilde{\mathbf{v}} \pm \tilde{ \mathbf{b}},$ which are the linear eigenmodes when $\beta=0$, i.e\ in ``pure'' MHD.  The physics at small $\beta$ is simplified in this basis because the nonlinearity in MHD turbulence is entirely generated by \emph{counter-propagating} Els\"asser populations; $\mathbf{z}^+$ and $\mathbf{z}^-$ interact with each other but not themselves. The Maxwell-Reynolds stress, which gives rise to momentum transport, is equivalent to the \emph{alignment} of counter-propagating Els\"assers: 
\begin{equation} 
\operatorname{Re} \langle \mathbf{z}^+_\mathbf{k} \cdot \mathbf{z}^-_\mathbf{-k} \rangle = |\tilde v_\mathbf{k}|^2 - |\tilde b_\mathbf{k}|^2.
\end{equation}
Similarly, cross-helicity is equivalent to an imbalance in Els\"asser populations; if we define $H_\mathbf{k} = \langle\tilde{ \mathbf{v}}_\mathbf{k} \cdot \tilde{ \mathbf{b}}_\mathbf{-k} \rangle,$ we have
\begin{equation}
\operatorname{Re} H_\mathbf{k} = \frac{1}{4}( |\mathbf{z}^+_\mathbf{k}|^2 - |\mathbf{z}^-_\mathbf{k}|.^2)
\end{equation}

In general, neither the Els\"asser basis nor the $\tilde{v}$, $\tilde{b}$ basis is preferred when the Rossby parameter is turned on. Instead, for finite $\beta$ the linear eigenmodes are Rossby-Alfv\'en modes, whose dispersion combines those of simple Rossby waves and Alfv\'en waves:
\begin{equation}
\omega= \frac{\omega_\beta \pm \sqrt{4\omega_A^2 + \omega_\beta^2}}{2},
\end{equation}
where $\omega_\beta$ and $\omega_A$ are respectively the Rossby and Alfv\'en frequencies. The coupling of Rossby-Alfv\'en waves of different species will mix the Els\"asser alignment and the cross-helicity, so that cross-helicity will play a role in all the spectral equations and help set the level of momentum transport in the system. This has the effect of severely complicating calculations.

In this work, we analytically study the momentum transport in this system, and in particular, its relationship with the cross-helicity. First, we show by a simple non-perturbative calculation---arguing on the basis of conservation of mean squared magnetic potential $\langle \tilde A^2 \rangle$---that the total cross-helicity in this system attains a finite, stationary value which scales with the turbulent magnetic energy. Second, we discuss a basic closure of the system within the framework of weak turbulence. Such an approximation should generally be reasonable when the mean field is especially large so that $\tilde v \ll b_0$. We derive the weak turbulence spectral equations for a generic system of interacting scalar fields and prove an identity that shows that for strong mean field, the stationary, real cross-helicity spectrum is set by the Els\"asser alignment spectrum. This identity also quantifies the balance between momentum transport and magnetic flux transport and can be linked with the Alfv\'enic character of the turbulence at small scales and its Rossby-like nature at large scales. We also carry out a calculation of the stationary spectra to leading order in $\beta$, with the proviso that this calculation only makes sense for sufficiently small parallel lengthscales. In the absence of $\beta$, there is no significant momentum transport. In contrast, we show that, when $\beta$ is turned on, variation over scales in the energy spectrum induces a cross-helicity shift at order $\beta$, which in turn results in an Els\"asser alignment, and thus momentum transport, at order $\beta^2$.

Next, we perform a set of simulations of this system for strong mean field and several values of $\beta$ and demonstrate that
\begin{itemize}
\item the simple non-perturbative calculation provides an accurate estimate of the actual stationary cross-helicity,
\item transition from Alfv\'enic to Rossby turbulence begins when the magnetic Rhines wavenumber exceeds the forcing wavenumber: $k_{\rm MR} = \sqrt{\beta/b_0} \gtrsim k_f^2$,
\item this transition is presaged by an increase in the global cross-helicity, which peaks around the critical $\beta$ where the above scales overlap,
\item the increase in cross-helicity is indeed linear in $\beta,$ and
\item the difference between kinetic and magnetic energies, a proxy for momentum transport, indeed increases as $\beta^2.$
\end{itemize}

We also present plots of the stationary spectra from simulation, which confirm the Rossby-like nature of the turbulence at large $\beta$ and the MHD-like nature at small $\beta$, and verify our predictions concerning the relationship between the cross-helicity and the Els\"asser alignment within weak turbulence as well as these quantities' dependence on $\beta$.

The remainder of this paper is organized as follows. In \S \ref{sec:th}, we calculate the stationary cross-helicity and discuss the weak turbulence closure of the spectral equations, including the relationship between the time-averaged cross-helicity spectrum and the time-averaged Els\"asser alignment spectrum. In \S \ref{sec:sim}, we present simulation results, both for global mean quantities and spectra. Finally, in \S \ref{sec:conc}, we summarize our results and discuss directions for future research.

\section{Theory}\label{sec:th}
\subsection{Stationary cross-helicity}\label{sec:zeld}
Our investigation begins with a simple, non-perturbative calculation of the cross-helicity at long times. The calculation is inspired by the Zel'dovich theorem \cite{zeld}; like that earlier result, here we use the conservation of mean-squared magnetic potential to relate large-scale transport properties at long times to small-scale dissipation. 

For the reader's convenience, we first rewrite the system (\ref{eq:mhd1})-(\ref{eq:mhd2}) with the mean magnetic field terms. We assume the system is forced kinetically only ($\tilde g = 0$):
\begin{align}
\label{eq:mhd3}\partial_t \nabla^2 \tilde \psi& + \beta \partial_x \tilde \psi = \{\tilde \psi, \nabla^2 \tilde \psi \} - \{\tilde A, \nabla^2 \tilde A \} + b_0 \partial_x \nabla^2 \tilde A + \nu \nabla^4  \tilde \psi + \tilde f \\
\label{eq:mhd4}\partial_t \tilde A& = \{ \tilde \psi, \tilde A\} + b_0 \partial_x \tilde \psi + \eta \nabla^2 \tilde A.
\end{align}

To be clear, $b_0$ represents an \emph{initial} mean field, and in principle the fluctuations $\tilde A$ could set up a mean field which opposes $b_0$; however, in this work we primarily consider weak turbulence, with $\langle \tilde b_x \rangle \ll b_0$ in practice.
 
Multiplying the second equation (\ref{eq:mhd4}) by $A$, integrating over space, and taking $t\to \infty$ yields 
\begin{equation} \frac12 \partial_t \langle \tilde A^2 \rangle = b_0 \langle \tilde A \partial_x \tilde \psi \rangle - \eta \langle (\nabla \tilde A)^2 \rangle =0 \end{equation}.
Note that $\langle \cdot \rangle$ is a global average, so that the triplet terms originating from the convective nonlinearity cancel (the nonlinearity conserves $\tilde A^2$). Thus the mean flux of magnetic potential is 
\begin{equation}
\label{eq:flux}
\langle \tilde A \partial_x \tilde \psi \rangle = \frac{\eta}{b_0} \langle\tilde b^2 \rangle.
\end{equation}
This is a consequence of Zel'dovich's theorem, which requires that the \emph{total} mean magnetic field must eventually decay in 2-D. The presence of the resistivity $\eta$ connects this decay to small-scale dissipation. However, the decay may be very slow \cite{pouquet78}. 

Next, we multiply the $A$ (\ref{eq:mhd4}) equation by $\nabla^2 \psi$ and the $\nabla^2 \psi$ equation (\ref{eq:mhd3}) by $A$; sum the resulting equations for $\tilde A \partial_t \nabla^2 \tilde \psi$ and $\tilde \nabla^2 \tilde \psi \partial_t \tilde A$; and finally integrate over space. This calculation is straightforward and yields
\begin{equation}
\partial_t \langle \tilde A \nabla^2  \tilde \psi \rangle = -\beta \langle \tilde A\partial_x \tilde \psi \rangle + (\eta + \nu) \langle \nabla^2 \tilde \psi \nabla^2 \tilde A \rangle + \langle \tilde A \tilde f \rangle.
\end{equation}
We then introduce characteristic lengthscales of the velocity and magnetic field variation and write $\langle \nabla^2 \psi \nabla^2 A \rangle \simeq \frac{1}{\ell_v \ell_b} \langle \mathbf{v} \cdot \mathbf{b} \rangle$. Finally, we substitute (\ref{eq:flux}) into the above, take the long-time limit (that is, we assume the cross-helicity is stationary), and obtain the estimate for the stationary cross-helicity (defined as $H\equiv \langle \tilde{\mathbf{v}} \cdot \tilde{\mathbf{b}} \rangle = - \langle \tilde A \nabla^2 \tilde \psi \rangle$)
\begin{equation}
H(t=\infty) = \frac{ \ell_v \ell_b}{\eta + \nu} \left(\frac{\beta \eta}{b_0}\langle \tilde b^2 \rangle - \langle \tilde A \tilde f \rangle \right).
\end{equation}
To good approximation we have $\langle \tilde A \tilde f \rangle\simeq 0$ , leading to the simpler expression
\begin{equation}
\label{eq:zeld}
H(t=\infty) = \frac{ \ell_v \ell_b}{1 + {\rm Pm}} \frac{\beta \langle\tilde  b^2 \rangle}{b_0},
\end{equation}
where ${\rm Pm} \equiv \nu / \eta$ is the magnetic Prandtl number. 

In practice, the forcing term $\langle \tilde A \tilde f \rangle$ will make a nonzero contribution to the total cross-helicity and cause it to drift randomly about its stationary mean value, and for small enough $\beta$ this drift can be significant (see App.~\ref{app:forcing_wwt}). If the system is magnetically forced, there will be an additional contribution from $\langle \nabla^2 \tilde \psi \tilde g \rangle$. 

When $\beta$ is not too large, $ \beta \lesssim b_0 k_f^2$ (as a reminder, $k_f$ is the scale at which the system is forced), we can estimate $\ell_v = \ell_b \simeq k_f^{-1}$, valid for weak turbulence. When $\beta$ is large, this is no longer a good estimate for $\ell_b$, as the magnetic field goes to small scales; a better estimate is then a magnetic Taylor microscale $\sqrt{\frac{\eta}{\varepsilon}\langle\tilde  b^2 \rangle}$, where $\varepsilon$ is again the energy injection rate (equivalent to the dissipation rate for stationary turbulence). 

Note that this calculation depended on the magnetic field being toroidally aligned; otherwise, conservation of $\tilde A^2$ is not enough to deduce the flux of magnetic potential across the planetary vorticity gradient. Note also that the cross-helicity could be very large in the case of a weak mean field and strong fluctuations.

Finally, it is important to point out that this argument did \emph{not} select the sign of the cross-helicity. We will see later that, for weak turbulence, the sign of the cross-helicity is determined by the imbalance between turbulent kinetic and magnetic energies.

\subsection{Weak turbulence closure of spectral equations}
We have derived an estimate for the stationary \emph{mean} cross-helicity; however, discerning transport properties requires studying the spectra. To this end, we seek a statistical closure which treats cross-correlations like the cross-helicity spectrum on an equal footing with autocorrelations like $\langle|\tilde v_\mathbf{k}|^2\rangle.$ The simplest approach, which we adopt here, is the weak turbulence theory of Sagdeev and Galeev \cite{sagdeev,zakharov}, though we note that \cite{grappin83} took an alternative approach (in pure MHD) based on the eddy-damped quasi-normal Markovian (EDQNM) closure theory. Formally a second-order time-dependent perturbation theory, weak turbulence assumes that the linear timescales are fast compared to nonlinear ones, so that the nonlinearity may be decomposed in terms of resonant triplet interactions between linear eigenmodes --- here, Rossby-Alfv\'en modes. For this approach to be valid, we must assume a strong mean magnetic field. Even here, the weak turbulence description will be dubious at large parallel lengthscales, where the linear frequencies vanish.

In $\mathbf{k}$-space, the $\beta$-plane MHD system is expressed as
\begin{align*}
&\partial_t \psi_{\mathbf{k}} + i \omega_{\beta,\mathbf{k}} \psi_\mathbf{k} + i \omega_{A,\mathbf{k}} A_\mathbf{k} = \\ &\quad\frac12 \int d^2\mathbf{k'} d^2\mathbf{k''} \, \delta(\mathbf{k}-\mathbf{k'}-\mathbf{k''}) \left[ (\mathbf{k'}\times \mathbf{k''}) \cdot \hat{z} \right]\frac{k'^2-k''^2}{k^2} \left(\psi_\mathbf{k'} \psi_\mathbf{k''} - A_\mathbf{k'} A_\mathbf{k''}\right) +  \frac{\tilde f_\mathbf{k}}{k^2}, \\
&\partial_t A_\mathbf{k} + i \omega_{A,\mathbf{k}} \psi_\mathbf{k} =  \frac12 \int d^2\mathbf{k'} d^2\mathbf{k''} \, \delta(\mathbf{k}-\mathbf{k'}-\mathbf{k''}) \left[(\mathbf{k'}\times \mathbf{k''}) \cdot \hat{z} \right]\left(A_\mathbf{k'} \psi_\mathbf{k''} - \psi_\mathbf{k'} A_\mathbf{k''} \right),
\end{align*}
where $\omega_{\beta,\mathbf{k}} = -\beta k_x/k^2$ is the Rossby frequency and $\omega_{A,\mathbf{k}} = b_0 k_x$ is the Alfv\'en frequency, and we have neglected forcing and dissipation terms. 

Assuming for simplicity that the linear damping rates $\eta k^2$ and $\nu k^2$ are much smaller than the real linear frequencies (or that ${\rm Pm} =1$),  the linear eigenmodes of this system are the two Rossby-Alfv\'en modes with frequencies
\[ \omega^\pm = \frac{\omega_\beta\pm\Omega}{2} \]
and corresponding amplitudes
\begin{equation}
\phi^\pm = \frac{1}{\Omega}(\omega^\pm \psi - \omega_A A),
\end{equation}
where $\Omega = \operatorname{sgn}(k_x) \sqrt{4 \omega_A^2 + \omega_\beta^2}$. As a reminder, $\omega_\beta = -\beta k_x/k^2$ and $\omega_A = k_x b_0$. The sign of the square root and the normalization have been chosen so that the $k_x \to 0$ limit is not problematic. In the $b_0\to \infty$ limit, these are just Els\"asser modes; as $b_0\to 0$, one mode becomes a Rossby wave and the other ceases to exist --- the amplitude and frequency simultanenously vanish. In this sense, the Rossby turbulence limit is singular. More precisely, the limits that must be taken are that of $b_0 k^2/\beta$, a dimensionless, lengthscale-dependent quantity, suggesting immediately that the turbulence should be Rossby-like when this parameter is small on typical scales and Alfv\'enic when it is large on typical scales. 

The classical weak turbulence theory, formulated for a single scalar field, is well-known \cite{sagdeev,zakharov}. However, its generalization to the case of multiple interacting scalar fields is rarely (if ever) seen in detail in the literature, so we present it here. Let $\{\phi^\alpha\}$ be a finite set of quadratically interacting linear eigenmodes described by the evolution equations
\begin{equation}
\label{eq:fieldeqs}
\partial_t \phi_\mathbf{k}^\alpha + i \omega^\alpha_\mathbf{k} \phi^\alpha_\mathbf{k} =\sum_{\beta \gamma} \frac12 \int d^2\mathbf{k'} d^2\mathbf{k''} \, \delta(\mathbf{k}-\mathbf{k'}-\mathbf{k''})  M^{\alpha \beta \gamma}_{\mathbf{k} ,\mathbf{k}',\mathbf{k}''} \phi^{\beta}_{\mathbf{k}'} \phi^{\gamma}_{\mathbf{k}''} + f^\alpha_\mathbf{k}.
\end{equation}
In particular, the linear frequency matrix is diagonal. The coupling coefficients are assumed without loss of generality to obey the symmetry condition $M^{\alpha \beta \gamma}_{\mathbf{k} ,\mathbf{k}',\mathbf{k}''}= M^{\alpha \gamma \beta}_{\mathbf{k} ,\mathbf{k}'',\mathbf{k}'}$. 

We define the correlators $C^{\alpha \alpha'}_\mathbf{k}$ by 
\[ \langle \phi^\alpha_\mathbf{k} \phi^{\alpha'}_{\mathbf{k}'} \rangle \equiv C^{\alpha \alpha'}_\mathbf{k} \delta(\mathbf{k}+ \mathbf{k}') e^{-i (\omega^\alpha_\mathbf{k} - \omega^{\alpha'}_\mathbf{k})t}.\]

After some work, it can be shown (see Appendix \ref{app:wwt}) that within the standard assumptions of weak turbulence (second-order perturbation theory, invoking both homogeneity and the random phase approximation), the correlators evolve according to the spectral equations
\begin{align}
\partial_t C^{\alpha \alpha'}_\mathbf{k} &= \sum_{\beta \gamma} \int d^2\mathbf{k'} d^2\mathbf{k''} \, \delta(\mathbf{k}-\mathbf{k'}-\mathbf{k''}) \bigg[ \pi |M^{\alpha \beta \gamma}_{\mathbf{k}, \mathbf{k}', \mathbf{k}''} |^2 C^{\beta \beta}_\mathbf{k'} C^{\gamma \gamma}_\mathbf{k''} \delta(\omega^\alpha_\mathbf{k} - \omega^\beta_\mathbf{k'} - \omega^\gamma_\mathbf{k''})\delta_{\alpha \alpha'} \nonumber \\ &+ M^{\alpha \beta \gamma}_{\mathbf{k}, \mathbf{k}', \mathbf{k}''} M^{\beta \alpha \gamma}_{\mathbf{k}', \mathbf{k}, -\mathbf{k}''} C^{\alpha \alpha'}_\mathbf{k} C^{\gamma \gamma}_\mathbf{k''} \left( \pi \delta(\omega^\alpha_\mathbf{k} - \omega^\beta_\mathbf{k'} - \omega^\gamma_\mathbf{k''}) + i \mathcal{ P} \frac{1}{\omega^{\alpha}_\mathbf{k} - \omega^\beta_\mathbf{k'} - \omega^\gamma_\mathbf{k''}}\right) \nonumber \\ &+ M^{\alpha' \beta \gamma*}_{\mathbf{k}, \mathbf{k}', \mathbf{k}''} M^{\beta \alpha' \gamma *}_{\mathbf{k}', \mathbf{k}, -\mathbf{k}''} C^{\alpha \alpha'}_\mathbf{k} C^{\gamma \gamma }_\mathbf{k''} \left( \pi \delta(\omega^{\alpha'}_\mathbf{k} - \omega^\beta_\mathbf{k'} - \omega^\gamma_\mathbf{k''}) - i \mathcal{ P} \frac{1}{\omega^{\alpha'}_\mathbf{k} - \omega^\beta_\mathbf{k'} - \omega^\gamma_\mathbf{k''}}\right)   \bigg] \\ &\quad + 2 \operatorname{Re} F^{\alpha \alpha'}_{\mathbf{k}}  \label{eq:coll}.
\end{align}
Here, $\mathcal{ P}$ means the Cauchy principal value is taken, and $\delta_{\alpha \alpha'}$ is a Kronecker delta. Note that in the case of a single mode with real coupling coefficients, the principal value terms cancel and we obtain the usual Sagdeev-Galeev theory. The Dirac deltas constrain the collision integral to resonances where the mismatch frequency $\Delta \omega = \omega^\alpha_\mathbf{k} - \omega^\beta_\mathbf{k'} - \omega^\gamma_\mathbf{k''}$ vanishes. The final term is due to the forcing and defined by the correlator $\langle f^\alpha_\mathbf{k}(t) f^{\alpha'}_\mathbf{-k}(t') \rangle = F^{\alpha \alpha'}_\mathbf{k} \delta(t-t')$ (see App.~\ref{app:forcing_wwt}).
t
To apply this result to our $\beta$-plane MHD system, where we have two scalar fields $\psi$ and $A$, we will will need to transform to the eigenbasis of two Rossby-Alfv\'en modes, $+$ and $-,$ and obtain the coupling coefficients in this basis. These are 
\begin{align}
M^{\pm++}_{\mathbf{k},\mathbf{k}' ,\mathbf{k}''}&= \frac{\hat{z}\cdot(\mathbf{k}'\times \mathbf{k}'')}{\Omega_\mathbf{k}} \left[\frac{k_x \omega^-_\mathbf{k'}}{ k_x'}-\frac{k_x \omega^-_\mathbf{k''}}{ k_x''} + \frac{k''^2-k'^2}{k^2}\omega^\pm_\mathbf{k}\left(1-\frac{\omega^-_\mathbf{k'} \omega^-_\mathbf{k''}}{k_x' k_x'' b_0^2}\right) \right] \label{eq:m1} \\
M^{\pm+-}_{\mathbf{k},\mathbf{k}' ,\mathbf{k}''} &=-\frac{\hat{z}\cdot(\mathbf{k}'\times \mathbf{k}'')}{\Omega_\mathbf{k}} \left[\frac{k_x \omega^-_\mathbf{k'}}{ k_x'} -\frac{k_x \omega^+_\mathbf{k''}}{ k_x''}+ \frac{k''^2-k'^2}{k^2}\omega^\pm_\mathbf{k}\left(1-\frac{\omega^-_\mathbf{k'} \omega^+_\mathbf{k''}}{k_x' k_x'' b_0^2}\right) \right] \label{eq:m2}\\
M^{\pm--}_{\mathbf{k},\mathbf{k}' ,\mathbf{k}''}&= \frac{\hat{z}\cdot(\mathbf{k}'\times \mathbf{k}'')}{\Omega_\mathbf{k}} \left[\frac{k_x \omega^+_\mathbf{k'}}{ k_x'}-\frac{k_x \omega^+_\mathbf{k''}}{ k_x''} + \frac{k''^2-k'^2}{k^2}\omega^\pm_\mathbf{k}\left(1-\frac{\omega^+_\mathbf{k'} \omega^+_\mathbf{k''}}{k_x' k_x'' b_0^2}\right) \right]. \label{eq:m3}
\end{align}

We also may express the Rossby-Alfv\'en correlators in the basis $E^K_\mathbf{k} = \langle |v_\mathbf{k}|^2 \rangle, E^M_\mathbf{k} = \langle |b_\mathbf{k}|^2 \rangle, H_\mathbf{k}:$  
\begin{align}
k^2 C^{\pm \pm}_\mathbf{k} &= \frac{1}{\Omega^2}\left(\omega_\pm^2 E^K_\mathbf{k} + \omega_A^2 E^M_\mathbf{k}- 2 \omega_A \omega_\pm \operatorname{Re} H_\mathbf{k}\right) \label{eq:cob1} \\
k^2 \operatorname{Re} (C^{+-}_\mathbf{k} e^{- i \Omega t}) &= -\frac{1}{\Omega^2}\left(\omega_A^2 (E^K_\mathbf{k} - E^M_\mathbf{k}) + \omega_\beta \omega_A \operatorname{Re} H_\mathbf{k} \right) \label{eq:cob2}  \\ 
k^2 \operatorname{Im} (C^{+-}_\mathbf{k} e^{- i \Omega t})&= -\frac{\omega_A}{\Omega} \operatorname{Im} H_\mathbf{k}. \label{eq:cob3} 
\end{align}

We see that the (real) cross-correlator between Rossby-Alfv\'en modes of opposite sign mixes the (real) cross-helicity and the Els\"asser alignment $E^K_\mathbf{k} - E^M_\mathbf{k},$ and naturally oscillates at frequency $\Omega= \omega^+- \omega^-.$

One may proceed by using the above expressions for the coupling coefficients, the transformation from the Rossby-Alfv\'en basis to the physical basis, and the inverse of that transformation to transform the spectral equations Eq.~\ref{eq:coll} into the physical basis. We will simplify the matter by computing low-order corrections in $\beta$, but first we present a useful identity.

\subsection{Cross-spectral identity}
Considering again the relationships between the spectra in the eigenbasis and in the physical basis --- Eqs.~(\ref{eq:cob1})-(\ref{eq:cob3}) --- one realizes that in a stationary state, the time-averaged Rossby-Alfv\'en cross-correlator must \emph{vanish} as long as $k_x\ne 0,$ since it naturally oscillates at the fast frequency $\Omega$. This imposes a major constraint on the stationary, time-averaged spectra: we must have
\begin{equation}
\label{eq:rel}
\langle E^K_\mathbf{k} - E^M_\mathbf{k} \rangle_t = \frac{\beta}{b_0 k^2} \operatorname{Re} \langle H_\mathbf{k} \rangle_t.
\end{equation}
($\langle \cdot \rangle_t$ is a time-average.) Thus, at the level of weak turbulence, a stationary cross-helicity will always be balanced by, and in a sense is \emph{equivalent} to, a deviation from energy equipartition. The latter also determines momentum transport. Another view is that the alignment of the velocity field with the magnetic field is determined by the alignment of the Els\"asser populations. Moreover, one may rearrange this to identify the left-hand side with the Lorentz force and the right-hand side with the turbulent emf and obtain:
\begin{equation}
\frac{\langle \partial_t \nabla^2 \tilde \psi \rangle_\mathbf{k}}{\langle \partial_t \tilde A\rangle_\mathbf{k}}  = k_{\rm MR}^2. 
\end{equation}
(The notations $\langle \partial_t \nabla^2 \tilde \psi \rangle_\mathbf{k}$ and $\langle \partial_t \tilde A\rangle_\mathbf{k}$ refer to the time-averaged RHS's of the evolution equations in Fourier space.)
Thus, the system organizes in such a way that, at each scale $k$, the ratio between the transport of momentum and the transport of magnetic field lines is (on average) $k_{\rm MR}^2/k^2$. This can be seen as a consequence of the fact that at large scales, the fluctuations are Rossby-like and carry momentum, and at small scales they are Alfv\'enic and do not transport momentum.

The vanishing of the cross-correlator also tells us that the imaginary part of the cross helicity vanishes, $\operatorname{Im} \langle H_\mathbf{k} \rangle_t = 0$. This is equivalent to the statement that the turbulent flux of $A$ vanishes in weak turbulence theory, and consistent with the Zel'dovich theorem for sufficiently large $b_0$ and/or sufficiently small fluctuations $\langle \tilde b^2 \rangle$. We should not be surprised, as even a \emph{weak} mean field is known to quench the turbulent resistivity \cite{cattaneo91,gruzinov94,gruzinov96,fan2019}.

This cross-spectral identity is simply the result of the linear mode structure of the problem. One expects a similar identity will hold for \emph{any} weakly turbulent system with multiple interacting modes. 

As an immediate application of the cross-spectral identity, we may integrate Eq.~\ref{eq:rel} over $\mathbf{k}$, taking care to properly treat the zonal modes ($k_x=0$), and substitute our expression for the total stationary cross-helicity Eq.~\ref{eq:zeld} into the RHS, and thusly obtain the estimate
\begin{equation}
\label{eq:dealf}
\frac{\langle \tilde v^2\rangle_{\rm NZ}}{\langle \tilde b^2\rangle} -1 \sim \frac{k_{\rm MR}^4}{(1+{\rm Pm})k_0^4},
\end{equation}
where $k_0$ is a characteristic lengthscale of the turbulence, and the subscript ``NZ'' means only nonzonal modes should be included. This relationship, which should hold for $\beta \lesssim \beta_c$ (the critical beta for transition to Rossby turbulence), helps quantify the degree that the turbulence has de-Alfv\'enized. $k_0$ may be crudely estimated by $k_f$.

\subsection{Pure MHD case}\label{sec:pure}
Let us first discuss the pure MHD system with $\beta=0.$  Here it is best to work in the Els\"asser basis. Let $E^\pm_\mathbf{k} = \langle |\mathbf{z}^\pm_\mathbf{k}|^2 \rangle$ and $P_\mathbf{k} = \langle \mathbf{z}^+_\mathbf{k} \cdot \mathbf{z}^-_\mathbf{k} \rangle$, where $\mathbf{z}^\pm = \mathbf{v} \pm \mathbf{b}$. The weak turbulence spectral equations are
\begin{align}
\partial_t E^\pm_\mathbf{k} &= \frac{\pi}{2} \int d^2 \mathbf{k}' d^2 \mathbf{k}'' \delta(\mathbf{k}-\mathbf{k}'-\mathbf{k}'') \delta(2b_0 k_x'') [\hat{z} \cdot (\mathbf{k}' \times \mathbf{k}'')]^2 \frac{(k^2+k'^2-k''^2)^2}{k^2 k'^2 k''^2}  E^\mp_\mathbf{k''} (E^\pm_\mathbf{k'}- E^\pm_\mathbf{k}) + \varepsilon_\mathbf{k} \label{eq:puremhd_e}\\
\partial_t \operatorname{Re} P_\mathbf{k} &= -\frac{1}{2} \int d^2 \mathbf{k}' d^2 \mathbf{k}'' \delta(\mathbf{k}-\mathbf{k}'-\mathbf{k}'')  [\hat{z} \cdot (\mathbf{k}' \times \mathbf{k}'')]^2 \left[\pi \delta(2 b_0 k_x'')  \operatorname{Re} P_\mathbf{k} - \frac{\cal P}{2 b_0 k_x''} \operatorname{Im} P_\mathbf{k} \right]  \nonumber  \\
&\quad \times \frac{k^2 +k'^2-k''^2}{k'^2 k''^2}  \left[ (k'^2+k''^2-k^2) E^+_\mathbf{k''} + (k^2+k'^2-k''^2) E^-_\mathbf{k''} \right] - 2b_0 k_x \operatorname{Im} P_\mathbf{k} + \varepsilon_\mathbf{k} \label{eq:puremhd_r}\\
\partial_t \operatorname{Im} P_\mathbf{k} &= -\frac{1}{2} \int d^2 \mathbf{k}' d^2 \mathbf{k}'' \delta(\mathbf{k}-\mathbf{k}'-\mathbf{k}'')  [\hat{z} \cdot (\mathbf{k}' \times \mathbf{k}'')]^2 \left[\pi \delta(2 b_0 k_x'')  \operatorname{Im} P_\mathbf{k} + \frac{\cal P}{2 b_0 k_x''} \operatorname{Re} P_\mathbf{k} \right]\nonumber \\
&\quad \times \frac{k^2 +k'^2-k''^2}{k'^2 k''^2}  \left[ (k'^2+k''^2-k^2) E^+_\mathbf{k''} + (k^2+k'^2-k''^2) E^-_\mathbf{k''} \right] + 2b_0 k_x \operatorname{Re} P_\mathbf{k}.  \label{eq:puremhd_i}
\end{align} 

Here, $\varepsilon_\mathbf{k} = 2\operatorname{Re} \langle \psi_\mathbf{k} f_\mathbf{-k} \rangle$ is the energy injection. The terms in Eqs.~(\ref{eq:puremhd_r}) and (\ref{eq:puremhd_i}) with coefficient $\pm 2 b_0 k_x$ arise due to the linear terms in the original field equations and can be recovered by restoring the oscillating factor $e^{-i \Omega t}$ to the cross-correlator $C^{+-}_\mathbf{k}.$ Let us refer to them as ``quasilinear'' terms.

The appearance of an energy injection term in Eq.~\ref{eq:puremhd_r} is the result of an imbalance between kinetic and magnetic forcing, which should be generic to most systems.

Now let us assume that the real cross helicity spectrum is identically zero, so that $E^+_\mathbf{k} =E^-_\mathbf{k}= E_\mathbf{k}$. If the forcing is of the form $\epsilon_\mathbf{k} \sim \delta(k-k_f)$, then the collision integral in Eq.~\ref{eq:puremhd_e} vanishes for $k_x=0$, whence one can easily compute $E(0,k_y)$ and then derive a functional equation for the energy spectrum $E_\mathbf{k}$. One finds that the spectrum will be supported only on the circles $k_x^2 + (k_y- n k_f)^2 = k_f^2,$ for $n \in \mathbb{Z}$.

If the cross-helicity spectrum is zero, then we can also rewrite Eqs.~(\ref{eq:puremhd_r}) and (\ref{eq:puremhd_i}) as the linear system
\begin{align}
\partial_t R_\mathbf{k} &= - \gamma_1 (\mathbf{k}) R_\mathbf{k} + (-2b_0 k_x + \gamma_2 (\mathbf{k}))I_\mathbf{k} + \varepsilon_\mathbf{k} \\
\partial_t I_\mathbf{k} &= - \gamma_1 (\mathbf{k}) I_\mathbf{k} - (-2b_0 k_x + \gamma_2 (\mathbf{k})) R_\mathbf{k}, 
\end{align}
where we have defined the shorthand $R_\mathbf{k} = \operatorname{Re} P_\mathbf{k}$ and $I_\mathbf{k} = \operatorname{Im} P_\mathbf{k}$, and where 
\be
\gamma_1 (\mathbf{k}) = \frac{\pi}{4b_0} \int d^2 \mathbf{k}' d^2 \mathbf{k}'' \delta(\mathbf{k}-\mathbf{k}'-\mathbf{k}'') \delta(k_x'') [\hat{z} \cdot (\mathbf{k}' \times \mathbf{k}'')]^2 \frac{k^2+k'^2-k''^2}{k''^2} E_\mathbf{k''}
\ee
and
\be
\gamma_2 (\mathbf{k}) = \frac{1}{4b_0} {\cal P}  \int d^2 \mathbf{k}' d^2 \mathbf{k}'' \delta(\mathbf{k}-\mathbf{k}'-\mathbf{k}'') \frac{1}{k_x''} [\hat{z} \cdot (\mathbf{k}' \times \mathbf{k}'')]^2 \frac{k^2+k'^2-k''^2}{k''^2} E_\mathbf{k''}.
\ee

$\gamma_2$ can be thought of as a renormalization of the ``bare'' frequency $2b_0 k_x$ by which $R$ and $I$ oscillate. It is easy to show that when $k_x$ is nonzero, $\gamma_1(\mathbf{k}) >0 $, as a consequence of the fact that $E_\mathbf{k}$ is nonnegative and not identically zero. ($\gamma_2(\mathbf{k})$, on the other hand, has no definite sign.) This means that, in the absence of the forcing, $R_\mathbf{k}$ and $I_\mathbf{k}$ would tend to zero as $t\to \infty$, for all $\mathbf{k}$, except when $k_x=0$ or $k_y=0$, and thus the magnetic and kinetic energy spectra would be asymptotically identical at finite wavenumbers. With $\varepsilon_\mathbf{k}$ turned on, we instead have the stationary spectra
\begin{align}
R_\mathbf{k} &= \frac{\gamma_1(\mathbf{k}) \varepsilon_\mathbf{k}}{\gamma_1(\mathbf{k})^2+(2b_0 k_x - \gamma_2(\mathbf{k}))^2} \\
I_\mathbf{k} &= \frac{(2b_0 k_x - \gamma_2(\mathbf{k})) \varepsilon_\mathbf{k}}{\gamma_1(\mathbf{k})^2+(2b_0 k_x - \gamma_2(\mathbf{k}))^2}.
\end{align}

To give an idea of the behavior of $\gamma_1$ and $\gamma_2$, we evaluate them for a couple simple energy spectra. First, with a flat spectrum $E_\mathbf{k} = C$ we have, to leading order in an ultraviolet cutoff $k_{\max},$:
\begin{align}
\gamma_1(\mathbf{k}) &= \frac{\pi C}{b_0} k_{\max} k_x^4 \\
\gamma_2(\mathbf{k}) &= -\frac{C}{2b_0} k_{\max}^2 k_x (4k_y^2 + \pi (k_x^2 -3k_y^2)).
\end{align}

Second, with a spectrum sharply peaked at a certain wavelength, $E_\mathbf{k} = \epsilon \delta(k-k_0),$ we have 
\begin{align}
\gamma_1(\mathbf{k}) &= \frac{\pi \epsilon}{b_0} k_x^2 k^2\\
\gamma_2(\mathbf{k}) &= \frac{\pi \epsilon}{4b_0} k_0 k_x (k_y^2-k_x^2).
\end{align}

We also estimate that for $i\in\{1,2\},$ $\gamma_i/k_x b_0 \sim \tilde v^2/b_0^2$ where $\tilde v$ is a typical fluctuation velocity of the turbulence; for weak turbulence this ratio should be small. Therefore we should have $R_\mathbf{k} \ll I_\mathbf{k} \simeq \varepsilon_\mathbf{k}/2 b_0 k_x$ at inertial ranges (though it need not be the case at large parallel scales). We also have $I_\mathbf{k}/E_\mathbf{k} \sim (\varepsilon_\mathbf{k}/ k_0 b_0)/(\varepsilon_\mathbf{k}/(\eta + \nu)k_0^2) \sim k_0(\eta+\nu) /b_0$ for a typical scale $k_0$, which should be small provided the mean field is large enough and/or the collisionality is not too strong. Thus we expect the ordering $R_\mathbf{k} \ll I_\mathbf{k} \ll E_\mathbf{k}. $ 

 It should be noted that for $k_x=0$, both $\gamma_1$ and $\gamma_2$ are zero, so at large parallel scales the damping is purely collisional and thus slow. We can then expect that the zonal $P_\mathbf{k}$ may be relatively large. 

Another important feature of the spectral equations (\ref{eq:puremhd_e})--(\ref{eq:puremhd_i}) is that, on account of the Alfv\'en resonance condition $k_x''=0$, only modes with equal $k_x$ can couple; nonlinear energy transfer only occurs in the $k_y$ direction \cite{tronko}. However, resonance broadening effects beyond the scope of weak turbulence are capable of mixing different $k_x$'s. 

\subsection{Small $\beta$ perturbation theory}\label{sec:perturb}

We now return to the general case $\beta\ne 0$. Our approach is small-$\beta$ perturbation theory about pure MHD ($\beta=0$) for the stationary spectra. We start with a pure MHD fixed point with no cross helicity and energy spectrum $E^+_\mathbf{k}=E^-_\mathbf{k} = E^{(0)}_\mathbf{k}$. Following the scaling arguments of the previous subsection, we will assume $P_\mathbf{k}$ to be negligible for this fixed point; it is true that nonzero $P_\mathbf{k}$ will generally make a contribution to the spectra that depends on $\beta$, but (in weak turbulence) the contributions purely from $E^{(0)}$ should be the largest. We will then turn on $\beta$ and perturbatively solve for the stationary spectra in powers of $\beta$ about this fixed point, in the spirit of mean-field electrodynamics. Our goal will be a lowest-order expression --- valid in the inertial range --- for $\operatorname{Re} P_\mathbf{k}$, to help us understand the momentum transport properties for stationary turbulence.

First, consider the Rossby-Alfv\'en wave-wave interactions that contribute to the collision integrals. In pure MHD, only counter-propagating Els\"asser populations interact, so the only processes with nonzero amplitude are $\pm, \mp \to \pm.$ The linear frequencies may be expanded as
\begin{equation} \omega_\pm = \pm \omega_A + \frac{\omega_\beta}{2} + O(\beta^2), \end{equation}
so the mismatch frequencies, which set the interaction time in weak turbulence, for these processes are 
\begin{equation}
\Delta \omega = \omega^\pm_\mathbf{k} - \omega^\pm_\mathbf{k'} - \omega^\mp_\mathbf{k''} \simeq \pm 2 b_0 k_x'' + \frac{1}{2} ( \omega_{\beta,\mathbf{k}} - \omega_{\beta,\mathbf{k'}} + \omega_{\beta,\mathbf{k}''}),
\end{equation}
where we have used the momentum conservation constraint $\mathbf{k'} + \mathbf{k''} = \mathbf{k}$. These processes present no particular issue for perturbation theory.

When $\beta$ is nonzero, previously forbidden processes now have nonzero amplitude. First, there is $\pm, \pm \to \pm$, with mismatch frequency
\begin{equation}
\Delta \omega \simeq \frac{1}{2} ( \omega_{\beta,\mathbf{k}} - \omega_{\beta,\mathbf{k'}} - \omega_{\beta,\mathbf{k}''}),
\end{equation}
equivalent to an effective Rossby-wave interaction. This frequency is $O(\beta)$, so the interaction is long-lived, $\tau \sim 1/|\beta|.$ These processes pick up a coefficient $M^2 \sim \beta^2$, so their overall contribution to the collision integrals is $O(\beta)$, still not necessarily problematic in perturbation theory. 

On the other hand the $\pm, \pm \to \mp$ processes have mismatch frequency 
\begin{equation}
\Delta \omega \simeq \pm 2b_0 k_x + \frac{1}{2} ( \omega_{\beta,\mathbf{k}} - \omega_{\beta,\mathbf{k'}} - \omega_{\beta,\mathbf{k}''}).
\end{equation}
One would like to make the formal expansion 
\begin{equation}
 \delta(\Delta \omega_0 + \beta \Delta \omega_1+ \dots) \simeq \delta(\Delta \omega_0) + \beta \Delta \omega_1 \delta'(\Delta \omega_0) +\dots, 
 \end{equation}
where $\delta'$ is the distributional derivative of the Dirac delta function, which may be handled in practice using integration by parts. This expansion is impossible for this process because the zeroth-order mismatch frequency $\Delta \omega_0 = \pm 2 b_0 k_x$ is independent of the integration variable $\mathbf{k}'$, and so the distributional derivative is undefined --- any attempt to perform the collision integral in this way will introduce a divergent factor $\left|\frac{\partial \Delta \omega_0}{\partial \mathbf{k}'}\right|^{-1}.$ Hence the contributions from this process must be handled more carefully.

An interpretation of this observation --- that na\"ive expansion of the resonance constraint fails --- is that all triplet Rossby-Alfv\'en processes \emph{except} $\pm, \pm \to \mp$ may be decomposed into interactions of pure Alfv\'en and/or Rossby waves. The basic features of the various kinds of Rossby-Alfv\'en triplet interactions are summarized in Table \ref{table}.

\begin{table}
\centering
\begin{tabular}{c|c|c|c}
Process & Permitted  & Notes for small $\beta$ & Contribution to spectra at low order \\
& in pure MHD? & & \\
\hline
$\pm, \mp \to \pm$ & Yes & Effective Alfv\'en-Alfv\'en & Shifts cross-helicity \\
& & interaction & \\
\hline
$\pm, \pm \to \pm$ & No & Effective Rossby-Rossby & Mixes energies in $\mathbf{k}$-space \\
& & interaction, long-lived & \\
\hline
$\pm, \pm \to \mp$ & No & Not decomposable into & Negligible, at least on small spatial scales\\
& & Rossby/Alfv\'en interactions & 
\end{tabular} 
\caption{Rossby-Alfv\'en triplet interactions in the $\beta$-plane system.}\label{table}
\end{table}

Let us now additionally assume the ordering $|k_x| \gtrsim k_{MR}$. Then the resonance curves for the $\pm, \pm \to \mp$ processes may be approximated by pairs of circles, parametrized by an angle $\theta\in [0,\pi)$:
\[
\mathbf{k'}=\left(\mp\frac{\beta}{4 b_0 k_x} \cos^2 \theta, \mp \frac{\beta}{4b_0 k_x} \cos \theta \sin \theta\right)
\]
and 
\[
\mathbf{k'} = \left(k_x\pm \frac{\beta}{4 b_0 k_x} \cos^2 \theta, k_y\pm \frac{\beta}{4b_0 k_x} \cos \theta \sin \theta\right).
\]

\begin{figure}
\centering
  \includegraphics[width=0.5\linewidth]{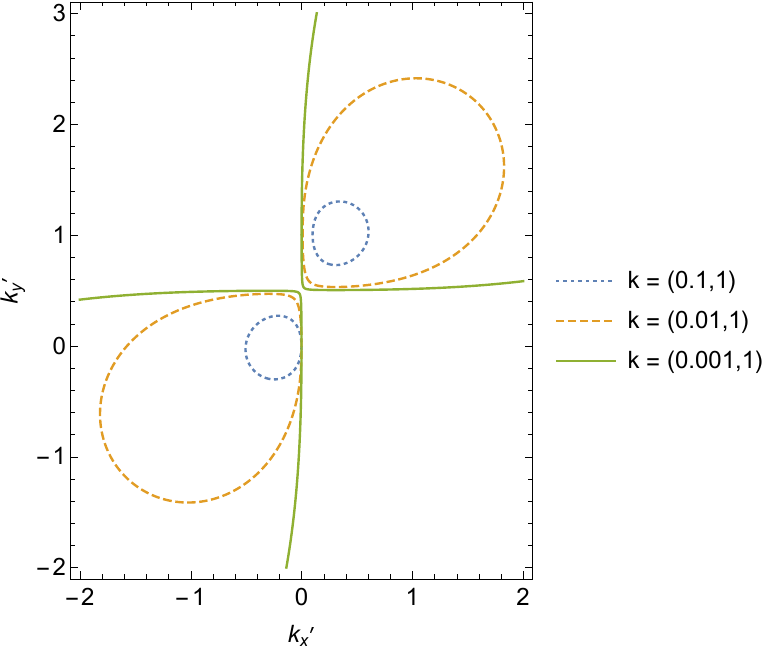}
\caption{Plots of the resonance curves for the Rossby-Alfv\'en process $-,- \to +$ at $k_y=1$, with $b_0=1$ and $\beta=0.2.$ Note that as $k_x$ becomes larger, the curves become well-approximated by circles.}\label{fig:resonance}
\end{figure}

Note that these curves will collapse to single points as $\beta\to 0$. This will correctly ensure that these processes do not contribute when $\beta \to 0$. Example resonance curves for this process are shown in Fig.~\ref{fig:resonance}. 

Restricting a collision integral to these curves will introduce a line element. For each curve we have
\[ \delta(\Delta \omega) d^2\mathbf{k}' \to \frac{1}{|\nabla_\mathbf{k'} \Delta \omega|} \left|\frac{d\mathbf{k'}}{ d \theta}\right| d\theta = \frac{\beta^2 \cos^2\theta} {32b_0^3 |k_x|^3} d \theta.\]

Thus, for this range of $k_x$, the contribution of the $\pm, \pm \to \mp$ processes to the spectra must be $O(\beta^3)$ or higher. The opposite case where $k_x$ is small requires techniques from strong turbulence theory to study effectively and is beyond the scope of this work.

Now let us proceed with the calculation of the stationary Els\"asser alignment $\operatorname{Re} P_k$. Using the cross-spectral identity, we can connect, at least to good approximation, $\operatorname{Re} P_k$ at a given order in $\beta$ to $E^+_\mathbf{k} - E^-_\mathbf{k}= 4 \operatorname{Re} H_\mathbf{k}$. Hence our strategy is to compute the cross helicity spectrum at lowest order in $\beta$, which we'll see is just linear order.

We make the ansatz
\begin{align}
\label{perturb}
E^\pm_\mathbf{k} &= E^{0}_\mathbf{k} + \beta E^{\pm,(1)}_\mathbf{k} + \beta^2 E^{\pm,(2)}_\mathbf{k} +O(\beta^3) \\
\operatorname{Re} P_\mathbf{k} &= \beta R^{(1)}_\mathbf{k} + \beta^2 R^{(2)}_\mathbf{k} +O(\beta^3) \\
\operatorname{Im} P_\mathbf{k} &= \beta I^{(1)}_\mathbf{k} + \beta^2 I^{(2)}_\mathbf{k}+ O(\beta^3)
\end{align}
and substitute it into the spectral equations (\ref{eq:coll}), using the coupling coefficients (\ref{eq:m1}--\ref{eq:m3}). We can then expand and solve order by order in $\beta.$

We will need the expansions
\begin{align}
E^+_\mathbf{k} &= 4k^2 C^{--}_\mathbf{k} + 2 \frac{\beta}{b_0} (\operatorname{Re} C^{+-}_\mathbf{k} - C^{--}_\mathbf{k}) + O(\beta^2) \\
E^-_\mathbf{k} &= 4k^2 C^{++}_\mathbf{k} - 2 \frac{\beta}{b_0} (\operatorname{Re} C^{+-}_\mathbf{k} - C^{++}_\mathbf{k}) + O(\beta^2) \\
\operatorname{Re} P_\mathbf{k} &= -4 k^2 \operatorname{Re} C^{+-}_\mathbf{k} + \frac{\beta}{b_0} (C^{--}_\mathbf{k}-C^{++}_\mathbf{k})+O(\beta^2) \\
\operatorname{Im} P_\mathbf{k} &= 4k^2 \operatorname{Im} C^{+-}_\mathbf{k} + O(\beta^2)
\end{align}

and their inverses
\begin{align}
k^2 C^{++}_\mathbf{k} &= \frac14 E^-_\mathbf{k} - \frac{\beta}{8b_0 k^2} (E^-_\mathbf{k} + \operatorname{Re} P_\mathbf{k}) +O(\beta^2) \\
k^2 C^{--}_\mathbf{k} &= \frac14 E^+_\mathbf{k} + \frac{\beta}{8b_0 k^2} (E^+_\mathbf{k} + \operatorname{Re} P_\mathbf{k}) +O(\beta^2) \\
k^2 \operatorname{Re} C^{+-}_\mathbf{k} &= -\frac14 \operatorname{Re} P_\mathbf{k} + \frac{\beta}{16b_0k^2} (E^+_\mathbf{k} - E^-_\mathbf{k}) + O(\beta^2) \\
k^2 \operatorname{Im} C^{+-}_\mathbf{k} &=\frac14 \operatorname{Im} P_\mathbf{k} +O(\beta^2).
\end{align}

The delta functions can be expanded in $\beta$ as well, except for the $\pm,\pm \to \mp$ processes. However, as already discussed, these processes can be ignored for the purposes of calculating spectra in the inertial range. In fact, we will only need the expansion of the delta function for the ``Alfv\'en-like'' process $\pm,\mp \to \pm$. One solves for the resonance condition at first order in $\beta$ and obtains 
\begin{equation}
\delta(\mathbf{k} - \mathbf{k}' -\mathbf{k}'') \delta( \omega^{\pm}_\mathbf{k} - \omega^{\pm}_\mathbf{k'} - \omega^{\mp}_\mathbf{k''})
= \delta(\mathbf{k} - \mathbf{k}' -\mathbf{k}'') \delta\left(\pm 2b_0 k_x''-\frac{\beta k_x}{2 k^2} +\frac{\beta k_x}{2 k'^2} \right).
\end{equation}
The terms contributed from this expansion represent the effect of dispersion induced by $\beta$, and they will depend on the spectral derivative $\partial_{k_x} E^{(0)}_\mathbf{k}.$

Using these expansions, obtaining the collision integrals at first order in $\beta$ is now a straightforward matter of tedious algebra, which is alleviated to some extent with the aid of a computer algebra system. Certain terms in the integrands will cancel due to symmetry under $\mathbf{k}' \leftrightarrow \mathbf{k''}$ exchange, and we can freely apply the constraint that $E^{(0)}_\mathbf{k}$ solves the zeroth order MHD equations (\ref{eq:puremhd_e})-(\ref{eq:puremhd_i}) to make some simplifications. One ultimately finds that the lowest-order effect of the Rossby-like $\pm, \pm \to \pm$ processes is to conservatively mix the total energy $E^+_\mathbf{k}+ E^-_\mathbf{k}$ among different $\mathbf{k}$ without affecting the balance $E^+_\mathbf{k}- E^-_\mathbf{k}$. Instead, only the Alfv\'en-like processes are found to contribute to the cross helicity. We find at last the first-order dynamics of the real cross helicity spectrum:
\begin{equation} 
\partial_t \operatorname{Re} H^{(1)}_\mathbf{k} = S_\mathbf{k} - \Gamma_\mathbf{k} \operatorname{Re} H^{(1)}_\mathbf{k} + \int d^2 \mathbf{k}' K(\mathbf{k},\mathbf{k'}) \operatorname{Re} H^{(1)}_\mathbf{k'} \label{eq:fredholm}.
\end{equation}
Here,
\begin{align}
S_\mathbf{k} &= \int d^2 \mathbf{k}' d^2 \mathbf{k''} \left[ \hat{z} \cdot (\mathbf{k}'\times \mathbf{k}'')\right]^2 \delta(\mathbf{k}-\mathbf{k'}-\mathbf{k''}) \delta (2 b_0 k_x'') \nonumber \\ &\quad \times \bigg\{ \frac{k_x^2 (k^2+k'^2-k''^2)(k'^2+k''^2-k^2)(k^2-k'^2)}{8b_0 k^4 k'^6 k''^2} E^{(0)}_\mathbf{k''}(E^{(0)}_\mathbf{k'}-E^{(0)}_\mathbf{k}) \nonumber \\ &\quad + \frac{k_x (k^2+k'^2-k''^2)^2(k^2-k'^2)}{16b_0 k^4 k'^4 k''^2} \left[ \partial_{k_x''} E^{(0)} _\mathbf{k''} (E^{(0)} _\mathbf{k'}-E^{(0)} _\mathbf{k}) - \partial_{k_x'} E^{(0)}_\mathbf{k'}E^{(0)} _\mathbf{k''} \right] \nonumber \\
&\quad + \frac{(k^2+k'^2-k''^2)^2(k^2-k'^2)}{16b_0 k^4 k'^4 k''^2} E^{(0)}_\mathbf{k} (E^{(0)}_\mathbf{k'} -E^{(0)}_\mathbf{k''}) \nonumber \\
&\quad + \frac{k^2+k'^2-k''^2}{4b_0 k^2 k'^2 k''^2} E^{(0)}_\mathbf{k'}  (E^{(0)}_\mathbf{k'} -E^{(0)}_\mathbf{k'}) \bigg\}, \\
\Gamma_\mathbf{k} &= \int d^2 \mathbf{k}' d^2 \mathbf{k''} \left[ \hat{z} \cdot (\mathbf{k}'\times \mathbf{k}'')\right]^2 \delta(\mathbf{k}-\mathbf{k'}-\mathbf{k''}) \delta (2 b_0 k_x'')  \frac{(k^2+k'^2-k''^2)^2}{2b_0 k^2 k'^2 k''^2} E^{(0)}_\mathbf{k''},  \\
K(\mathbf{k},\mathbf{k'}) &=  \frac{1}{b_0^2 k^2 k'^2 (\mathbf{k}-\mathbf{k'})^2 } \left[ \hat{z} \cdot (\mathbf{k}\times \mathbf{k}')\right]^2 \bigg[\delta(k_x-k_x')  (\mathbf{k} \cdot \mathbf{k'})^2 E^{(0)}_{\mathbf{k}-\mathbf{k'}} \nonumber \\ &\quad + \delta(k_x')  (k^2 -\mathbf{k}\cdot \mathbf{k'}))^2 (E^{(0)}_{\mathbf{k}}-E^{(0)}_{\mathbf{k}-\mathbf{k'}})\bigg].
\end{align}
Note in particular that the source term $S_\mathbf{k}$ vanishes if $E_\mathbf{k}^{(0)}={\rm const.}$, so the generation of cross-helicity is due to variations in the energy spectrum across scales. After imposing stationarity --- $\partial_t \operatorname{Re} H^{(1)}_\mathbf{k} =0$ --- Eq.~\ref{eq:fredholm} becomes a Fredholm integral equation of the second kind for $\operatorname{Re} H_\mathbf{k}^{(1)}.$ We can (at least formally) express the solution as a Liouville-Neumann series, i.e.\
\begin{equation} 
\operatorname{Re} H^{(1)}_\mathbf{k} = \frac{S_\mathbf{k}}{\Gamma_\mathbf{k}} + \int d^2 \mathbf{k}' \frac{K(\mathbf{k},\mathbf{k'})}{\Gamma_\mathbf{k'}} S_\mathbf{k'} + \int d^2 \mathbf{k}' d^2 \mathbf{k''} \frac{K(\mathbf{k},\mathbf{k'})K(\mathbf{k'},\mathbf{k''})}{\Gamma_\mathbf{k'}\Gamma_\mathbf{k''}} S_\mathbf{k''} + \dots.
\end{equation}

Finally, we compute the stationary Els\'asser alignment spectrum. At first order, it is is vanishing: we have
\begin{align}
\partial_t R^{(1)}_\mathbf{k} &= -\int d^2 \mathbf{k'} d^2 \mathbf{k''} \, [\hat{z} \cdot (\mathbf{k'} \times \mathbf{k''}) ]^2 \delta(\mathbf{k} - \mathbf{k}' -\mathbf{k}'') \nonumber \\ &\quad \times \frac{(k^2+k'^2-k''^2)}{2k^2 k''^2} E^{(0)}_\mathbf{k''} \bigg[ R^{(1)}_\mathbf{k} \delta(2 b_0 k_x'') - \mathcal{P}\frac{ I^{(1)}_\mathbf{k}}{2b_0 k_x''}\bigg] - 2b_0 k_x I^{(1)}_\mathbf{k} \\
\partial_t I^{(1)}_\mathbf{k} &= - \int d^2 \mathbf{k'} d^2 \mathbf{k''} \, [\hat{z} \cdot (\mathbf{k'} \times \mathbf{k''}) ]^2 \delta(\mathbf{k} - \mathbf{k}' -\mathbf{k}'') \nonumber \\&\quad\times \frac{(k^2+k'^2-k''^2)}{2k^2k''^2} E^{(0)}_\mathbf{k''} \bigg[ I^{(1)}_\mathbf{k} \delta(2 b_0 k_x'') + \mathcal{P}\frac{ R^{(1)}_\mathbf{k}}{2b_0 k_x''}\bigg] + 2b_0 k_x R^{(1)}_\mathbf{k},
\end{align}
and using similar arguments as in Sec.~\ref{sec:pure}, we conclude that $R^{(1)}_\mathbf{k}$ and $I^{(1)}_\mathbf{k}$ should both tend to zero.

At second order, the imaginary part of the Els\"asser alignment (which is associated with the flux of $A$) is controlled by the equation
\be
\partial_t I^{(2)}_\mathbf{k} = -2b_0 k_x R^{(2)}_\mathbf{k} + \frac{ 2k_x}{k^2} \operatorname{Re} H^{(1)}_\mathbf{k} + \left(\partial_t I^{(2)}_\mathbf{k}\right)_{\rm coll.},
\ee
where $\left(\partial_t I^{(2)}_\mathbf{k}\right)_{\rm coll.}$ are the nonlinear terms coming from the collision integral. We contend that for weak turbulence, the dominant balance is between the quasilinear terms, so that 
\begin{equation}
R^{(2)}_\mathbf{k} = \frac{1}{b_0 k_x^2} \operatorname{Re} H^{(1)}_\mathbf{k}.
\end{equation}
This is just a restatement of the cross-spectral identity, which we now see is equivalent to balance between the quasilinear terms in the dynamics of $\operatorname{Im} P_\mathbf{k}.$ We conclude that the leading order behavior of the Els\"asser alignment spectrum is $O(\beta^2).$

Let us now summarize what we have found. When $\beta$ is turned on, effects due to non-uniformity of the energy spectrum, including some related to the dispersion in the mode frequencies induced by $\beta$, result in a shift in the (real) cross-helicity proportional to $\beta$. This, in turn, is balanced by an Els\"asser aligment shift that is quadratic in $\beta$. (These results are not valid, however, for the spectra along $k_x=0$, which are stationary in weak turbulence.) We can therefore predict that, for stationary turbulence (and $\beta$ not too large), the total cross helicity $\langle H\rangle \propto \beta$ and the nonzonal deviation from equipartition is $\langle \tilde v^2_{\rm NZ}\rangle -\langle \tilde b^2\rangle  \propto  \beta^2$ (plus a small constant due to the forcing), which we confirm in the following section.


\section{Simulation results}\label{sec:sim}
\subsection{Stationary mean energies and cross-helicity}
We now present simulation results for this system, on a periodic box of size $2 \pi \times 2 \pi$ with a $512 \times 512$ mesh. We used a pseudospectral code within the Dedalus framework \cite{dedalus}, timestepped with a third-order, 4-stage DIRK+ERK scheme \cite{ascher1997}. The vorticity was forced in an annulus in Fourier space, centered at $k_f = 32$ and with a width of 8. The forcing had a fixed energy injection rate of $\varepsilon = 10^{-3}$ and a correlation time of $\tau_c = 2 \times 10^{-3}$. The choice of box size and correlation time can be thought of fixing a nondimensionalization of the equations. All simulations had fixed $\nu = \eta = 10^{-4}$ and $b_0 = 2$, and $\beta$ was varied over a broad range. Hyperviscous ($\nabla^4$) terms with coefficients $10^{-8}$ were included in each equation to improve the stability properties. For these simulations, the magnetic Reynolds number at the forcing scale ranged from $36 \le\mathrm{Rm}_f \le 211$, with the smallest numbers corresponding to the largest $\beta$. The simulations were initialized with a small random vorticity seed and then evolved to reach a state wherein macroscopic variables of interest (including turbulent energies, cross-helicity, mean-squared magnetic potential, and enstrophy) have stabilized. Within this state, the kinetic and magnetic energies, cross-helicity, and zonal mean (kinetic) energy fraction ${\rm zmf} = \int dy (\int dx \, \tilde v_x)^2 /(2 \pi \int dx dy  \, \tilde v^2)$ were all averaged over about ten diffusion times $10 \tau_D = 10(\nu k_f^2)^{-1}\simeq100.$ A zmf exceeding $\sim 0.5$ signals an inverse cascade --- most of the kinetic energy is at large parallel scales --- and is indicative of a transition to Rossby turbulence. 

\begin{figure}
\centering
\includegraphics[width=\columnwidth]{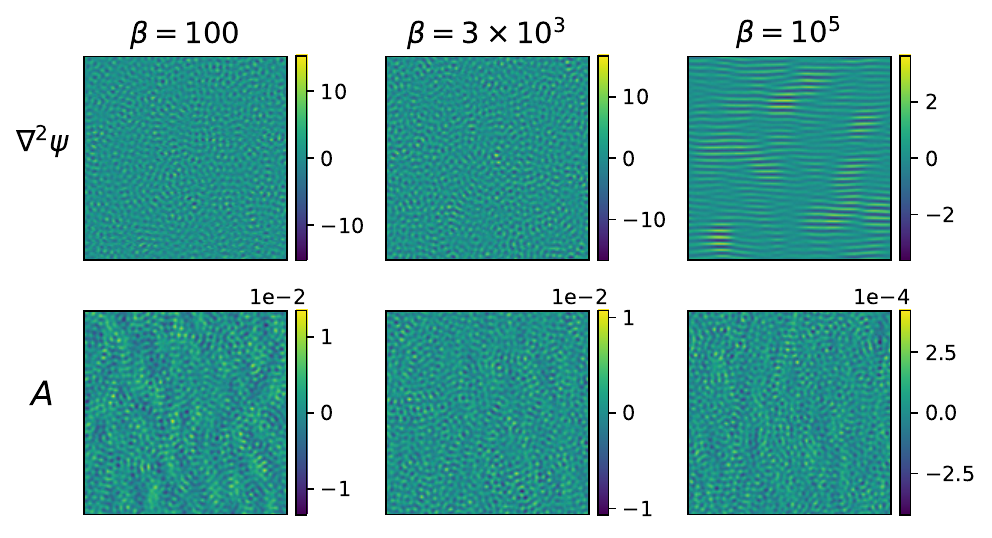}
\caption{Simulation snapshots at $t=150$, for $\beta=100,3\times10^3,$ and $10^5$. The first row shows the vorticity $\nabla^2 \tilde \psi$ as a function of position; the second row shows the magnetic potential $\tilde A$. Each column shows the results for a different value of $\beta$.} \label{fig:snapshots}
\end{figure}

\begin{figure}
\centering
\includegraphics[width=\columnwidth]{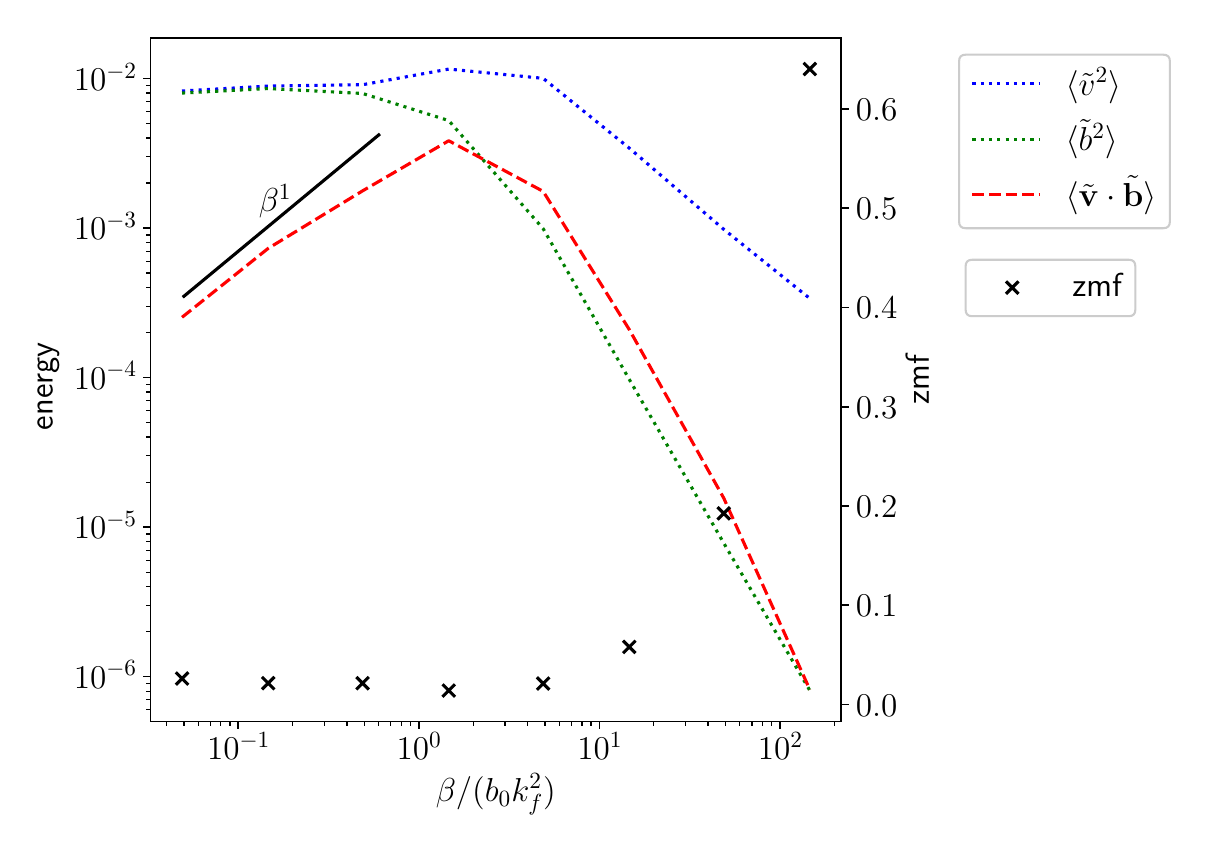}
\caption[Plot of stationary turbulent energies, cross-helicity, and zmf.]{Plot of stationary turbulent energies, cross-helicity, and zmf obtained from simulation. All simulations were kinetically forced at an energy injection rate $\varepsilon =10^{-3}$ at a scale $k_f = 32$ and had a mean field $\mathbf{b}_0 = 2 \hat{x}.$ The transition from Alfv\'enic to Rossby turbulence, signified by an imbalance between kinetic and magnetic energies and a large zmf, appears to start around a critical value $\beta = b_0 k_f^2$. This critical value, in general, will also depend on the resistivity. The transition is presaged by an increase in the mean cross helicity, which peaks around the critical value. $\beta$ is normalized in the plot by the critical value. The zmf becomes large ($>0.5$) when $\beta\gg b_0 k_f^2,$ signaling an inverse cascade. The cross-helicity is roughly linear with $\beta$ for small $\beta$, in agreement with theory.} \label{fig:chplot-1}
\end{figure}

\begin{figure}
\centering
\includegraphics[width=\columnwidth]{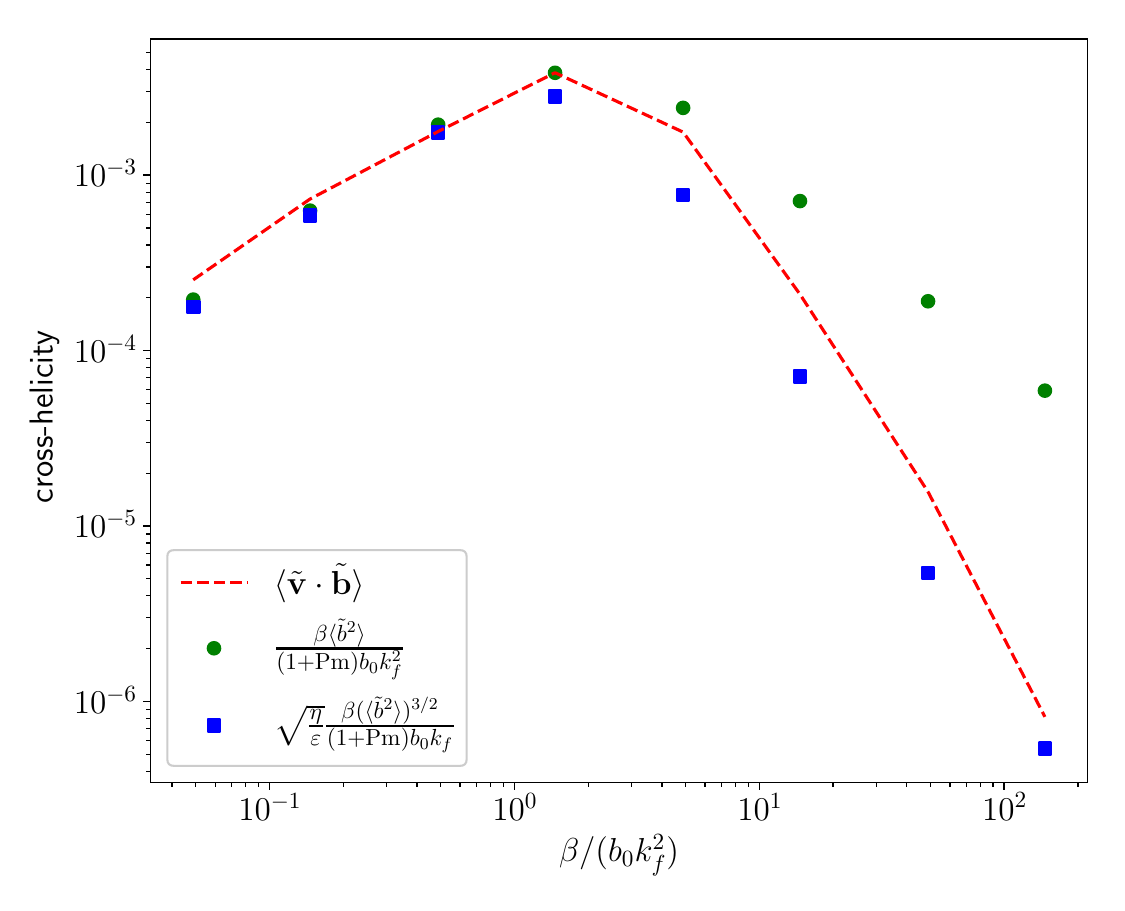}
\caption[Plot of stationary cross-helicity from simulation, compared with analytical predictions.]{We again plot the cross-helicity from simulation (red dashed line) and compare it to the prediction from \S~\ref{sec:zeld}. We use two estimates for the lengthscales: using $\ell_b = \ell_v = k_f^{-1}$ (green circles), which ceases to be good as $\beta$ is large since the magnetic fluctuation scale becomes small, and a second estimate (blue squares) using a magnetic Taylor microscale $ \ell_b = \sqrt{\eta/\varepsilon \langle \tilde b^2 \rangle}.$} \label{fig:chplot-2}
\end{figure}

Snapshots from the end of the simulations ($t=150$) are shown in Fig.~\ref{fig:snapshots}, for three values of $\beta$. For the largest $\beta$, zonal flows are clearly present. The results for the stationary energies and cross-helicity are plotted in Fig.~\ref{fig:chplot-1}; in Fig.~\ref{fig:chplot-2}, the cross-helicity is compared to the predictions of \S~\ref{sec:zeld}.

There is a wealth of information in these plots and it is worth discussing. When $\beta < b_0 k_f^2,$ the magnetic and kinetic energies are nearly equal, suggesting that the turbulence is Alfv\'enized. As $\beta$ grows larger, this equipartition begins to break down, and the magnetic energy drops, signaling the possible generation of a Reynolds stress. When $\beta$ is significantly greater than $b_0 k_f^2,$ most of the energy is in the flow and concentrated at large parallel lengthscales, signaling an inverse cascade and a transition to Rossby turbulence. 

Meanwhile, as $\beta$ nears the transition point, the cross-helicity grows, peaks near $\beta = b_0 k_f^2$, and then drops again. The growth of the cross-helicity appears to be linear in $\beta$ for small $\beta$, in agreement with the conclusions of Sec.~\ref{sec:perturb}. An estimate for the stationary cross-helicity using $\ell_b = \ell_v = k_f^{-1}$ begins to break down for large $\beta$; in this regime, the estimate $\ell_b = \sqrt{\frac{\eta}{\varepsilon}\langle\tilde b^2 \rangle}$ better fits the data.

\subsection{Flux of magnetic potential}
\begin{figure}
\centering
\includegraphics[width=\columnwidth]{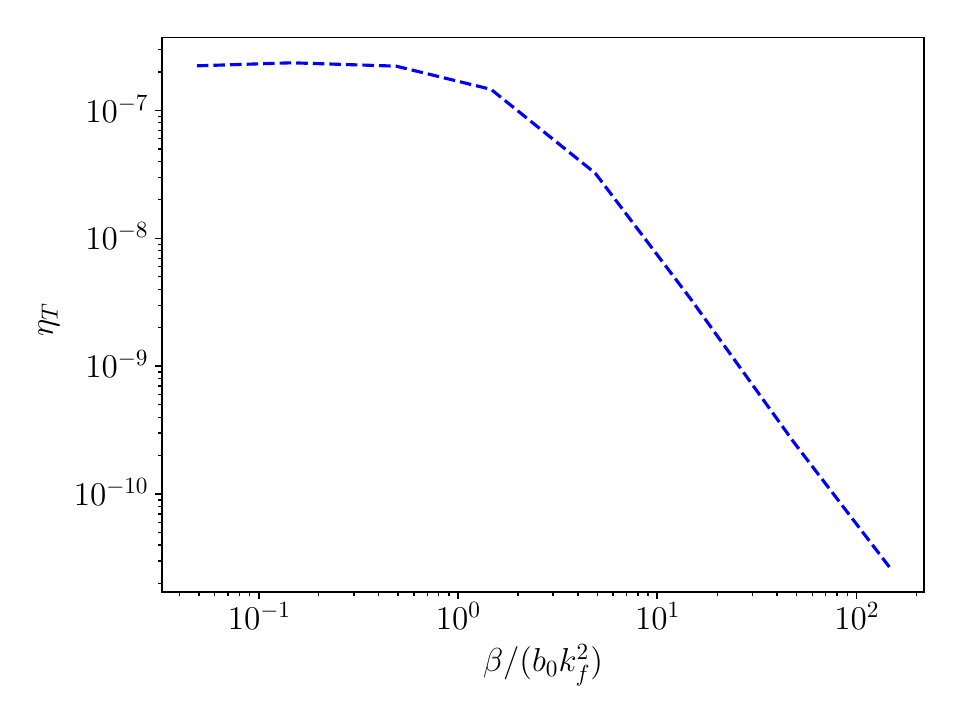}
\caption{Plot of the turbulent resistivity $\eta_T= -\langle \tilde v_y \tilde A \rangle/b_0$ as a function of $\beta.$}
\label{fig:flux}
\end{figure}

In Fig.~\ref{fig:flux} we plot the mean turbulent resistivity $\eta_T= -\langle \tilde v_y \tilde A \rangle/b_0$ as a function of $\beta$. It is generally very small, $\eta_T < \eta$, having been quenched by the strong mean field. This is consistent with the weak turbulence expectation $\operatorname{Im} H_\mathbf{k} = 0.$ It drops precipitously when $\beta > b_0 k_f^2$ due to the lack of magnetic activity --- the rms turbulent magnetic potential $\langle \tilde A^2 \rangle^{1/2}$ becomes very small. We also note that the instantaneous flux oscillates on fast timescales, with an amplitude roughly an order of magnitude larger than the mean flux.

\subsection{(De-)Alfv\'enization}
\begin{figure}
\centering
\includegraphics[width=\columnwidth]{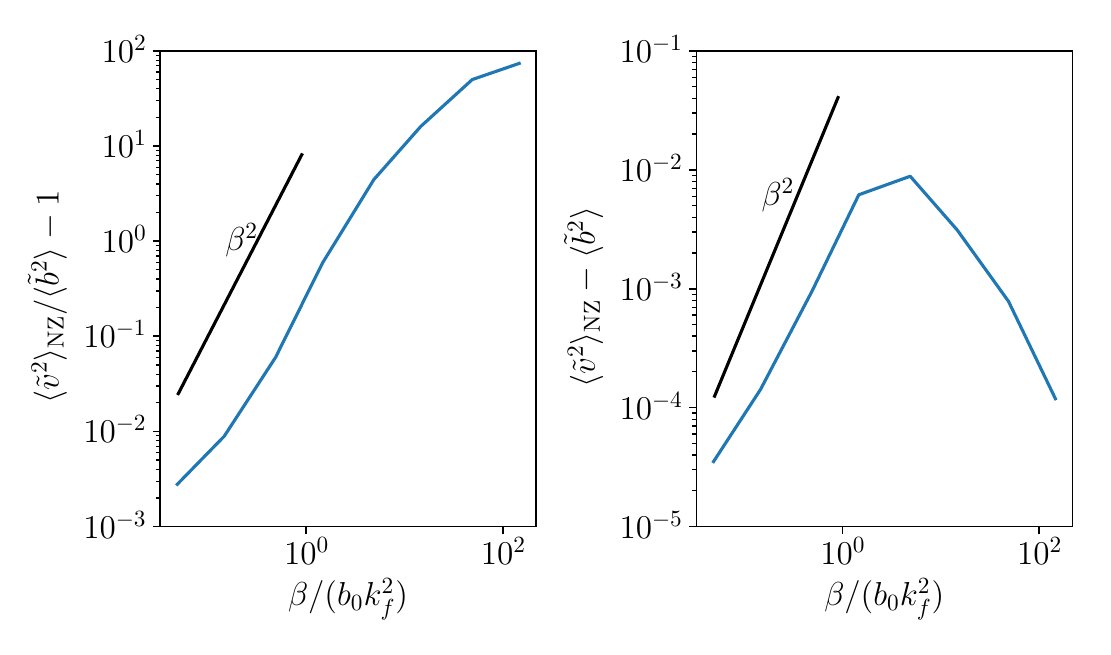}
\caption{Plots of $\frac{\langle \tilde v^2\rangle_{\rm NZ}}{\langle \tilde b^2\rangle} -1$ and $\langle \tilde v^2\rangle_{\rm NZ}-\langle \tilde b^2\rangle$ compared to $\beta^2$ scaling predicted by our theory.}
\label{fig:dealf}
\end{figure}

Equation \ref{eq:dealf} predicts that, for sufficiently small $\beta$, the quantity $\frac{\langle \tilde v^2\rangle_{\rm NZ}}{\langle \tilde b^2\rangle} -1$, which measures the degree to which the turbulence is (de-)Alfv\'enized, should scale as $\beta^2$. Similarly, we showed in Sec.~\ref{sec:perturb} that the energy differential itself, $\langle \tilde v^2\rangle_{\rm NZ}-\langle \tilde b^2\rangle$, should have the same scaling. These expectations are confirmed in Fig.~\ref{fig:dealf}, with good agreement for $\beta \lesssim b_0 k_f^2.$ In the Rossby regime, this estimate breaks down because large parallel lengthscales become very important, and the weak turbulence assumption that the dominant linear timescales are fast is no longer valid. 

\subsection{Spectra}
\begin{figure}[htp]
\centering
\includegraphics[width=\columnwidth]{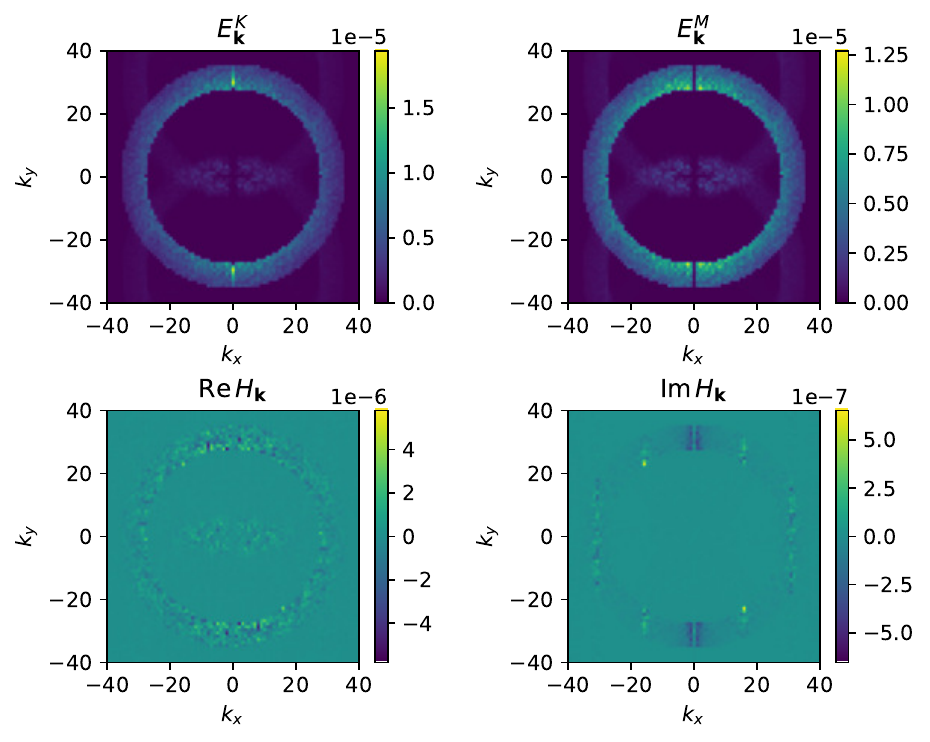}
\caption{Time-averaged spectra for $\beta=0$ (pure MHD).}
\label{fig:spectra0}
\end{figure}

\begin{figure}[htp]
\centering
\includegraphics[width=\columnwidth]{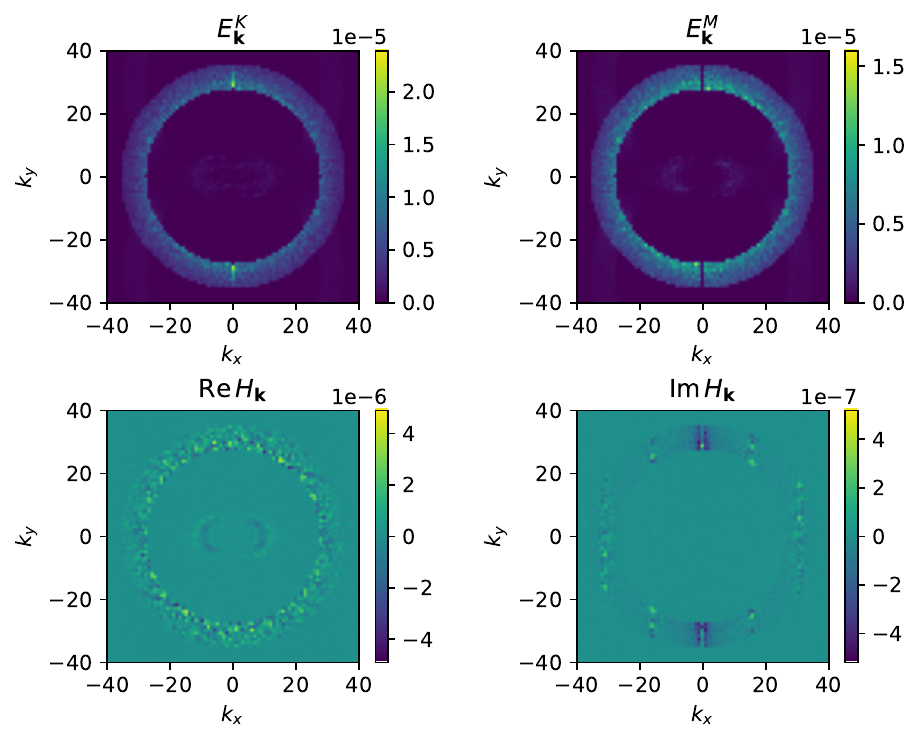}
\caption{Time-averaged spectra for $\beta=100,$ corresponding to the small-$\beta$ regime.}\label{fig:spectra1e2}
\end{figure}

\begin{figure}[htp]
\centering
\includegraphics[width=\columnwidth]{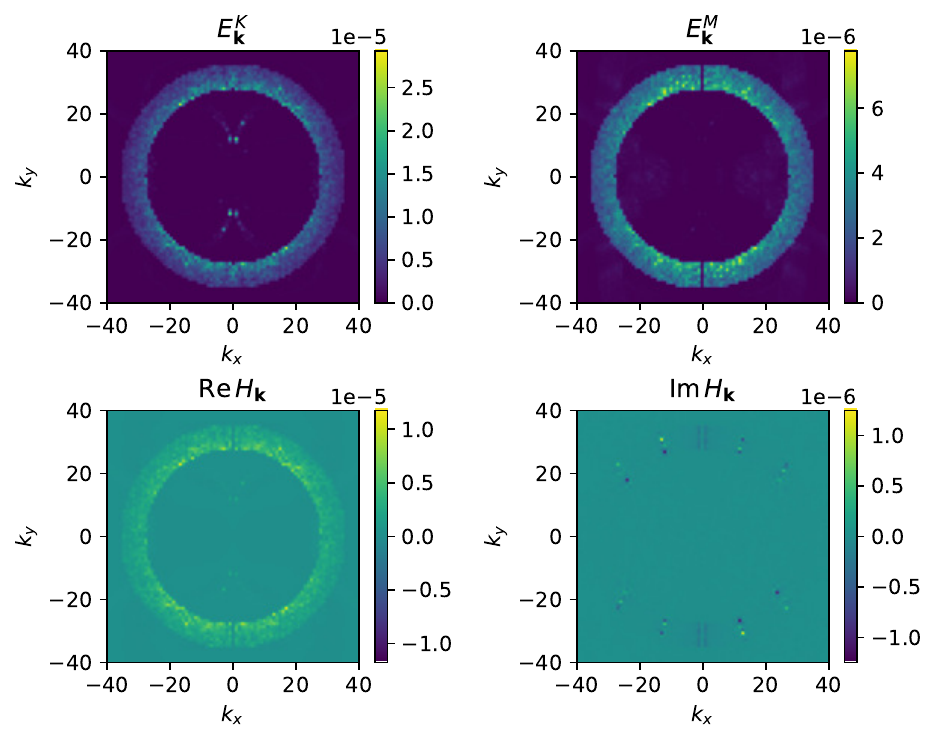}
\caption{Time-averaged spectra for $\beta=3\times10^3$, corresponding to the intermediate-$\beta$ regime.}\label{fig:spectra3e3}
\end{figure}

\begin{figure}[htp]
\centering
\includegraphics[width=\columnwidth]{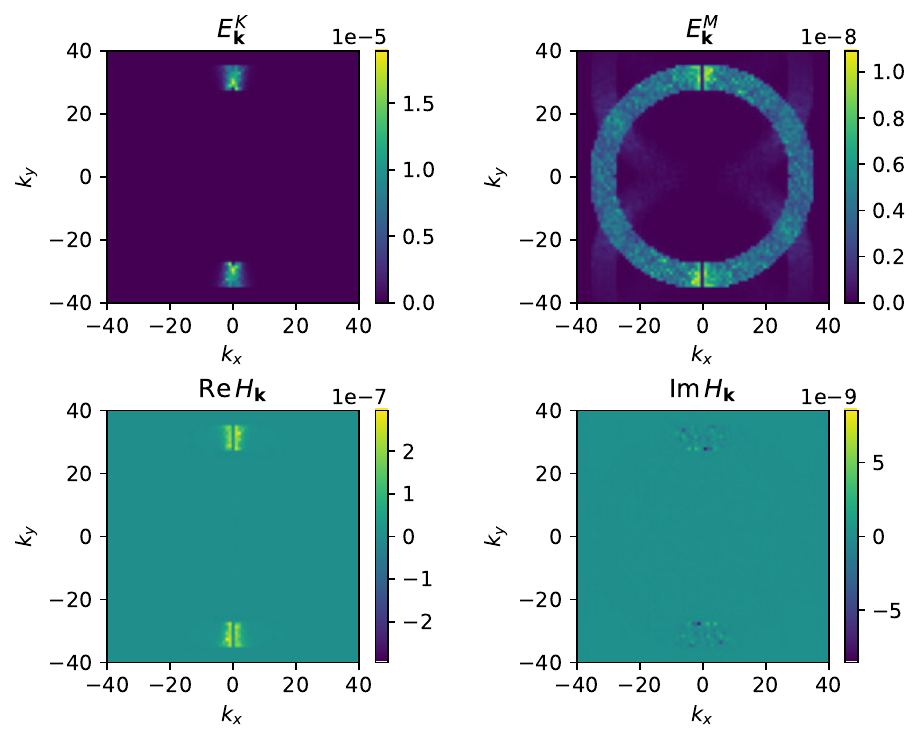}
\caption{Time-averaged spectra for $\beta=1\times10^5$, corresponding to the large-$\beta$ regime. }\label{fig:spectra1e5}
\end{figure}

We now turn our attention to the spectra. In Figs.~\ref{fig:spectra0}--\ref{fig:spectra1e5}, we plot the spectra, averaged over the same time window as in the previous subsection, for $\beta=0,100,3\times 10^3,$ and $10^5$, respectively representing the pure MHD case, the small-$\beta$ (Alfv\'enic) case, the intermediate-$\beta$ (transitional) case, and the large-$\beta$ (Rossby-like) case. The plots are shown over the most important range which includes the circle $k=k_f$. 

Let us begin with the pure MHD case. The energy is mostly confined to the forcing annulus, and consistent with the claims of Sec.~\ref{sec:pure}, the rest of the energy is mostly on the circles $k_x^2 + (k_y \pm k_f)^2 =k_f^2$. The kinetic and magnetic energy spectra are very similar, as one expects; the primary difference is that the kinetic spectra have a significant zonal component. The real-cross helicity spectrum is relatively small and appears the primary contribution is random. We expect that if the time-averaging window were extended, the real cross-helicity would be even smaller as the random component would be suppressed. Finally, the imaginary cross-helicity shows clear structure and is mostly confined to a few locations on the $k=k_f$ circle. This is also consistent with Sec.~\ref{sec:pure}: we roughly expect $I_\mathbf{k}$ to be largest in magnitude where $\gamma_2/k_x$ vanishes or is very large. In the simple examples of $\epsilon_\mathbf{k}$ we considered, $\gamma_2/k_x$ vanishes on a pair of lines of the form $k_x = \pm a k_y$, and it is large when $k_x$ is small and $k_y$ is large. When we further constrain $k\simeq k_f$, we should indeed expect a result similar to that in Fig.~\ref{fig:spectra0}.

Next, when $\beta=100$,  the change in the energy spectra relative to $\beta=0$ is slight; the primary differences are that the circles $k_x^2 + (k_y \pm k_f)^2 =k_f^2$ have been partially destroyed, and the kinetic energy is generally slightly larger in magnitude than its magnetic counterpart. The imaginary cross-helicity is also very similar to that of the pure MHD case. However, in the real cross-helicity, we now see some structure, particularly at small $k$. We suggest these are the result of non-uniformity in the energy spectra as previously argued.

When $\beta$ increases to $3\times 10^3$ (slightly more that $b_0 k_f^2$), the  difference in magnitude between $E^K$ and $E^M$ becomes much larger. We also see the appearance of interesting new features in the kinetic spectra at small but finite $k_x$, evidently a higher-order nonlinear effect not captured by our theory. The real cross-helicity is now significantly larger and has an obvious sign, as dictated by quasilinear balance.

Finally, when $\beta=10^5$, the kinetic energy is almost entirely concentrated near $k_y=\pm k_f$ and $k_x=0$, consistent with the condensation to large scales and the development of zonal flows. The real cross-helicity has accordingly collapsed to the same locations. The magnetic energy and imaginary cross-helicity are almost negligibly small in magnitude, as Alfv\'en wave activity is virtually nonexistent.

\begin{figure}[htp]
\centering
\includegraphics[width=\columnwidth]{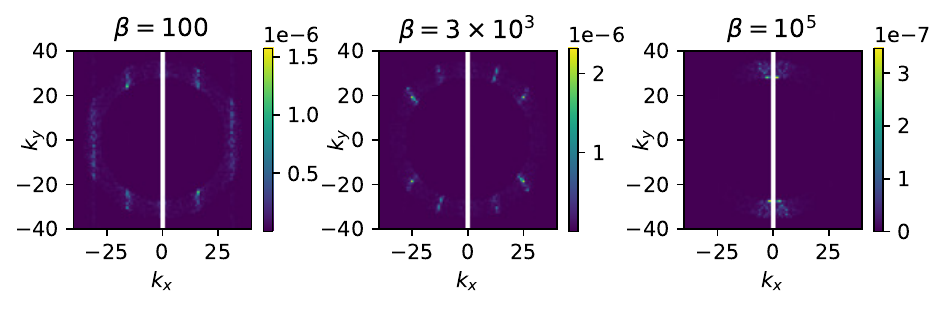}
\caption{Plots of the absolute difference between the time-averaged kinetic energy spectra and the weak turbulence estimate, i.e.\ $|E^K_\mathbf{k} - E^M_\mathbf{k} - \frac{\beta}{b_0 k^2} \operatorname{Re} H_\mathbf{k}|.$ We have excluded the irrelevant $k_x=0$ slice. Compared to typical values for $E^K_\mathbf{k}$ in Figs.~\ref{fig:spectra1e2}--\ref{fig:spectra1e5}, the error is generally small. These plots also highlight the $\mathbf{k}$ where nonlinear effects are important.}\label{fig:compare}
\end{figure}

In Fig.~\ref{fig:compare}, we directly evaluate the cross-spectral identity by plotting the residual $|E^K_\mathbf{k} - E^M_\mathbf{k} - \frac{\beta}{b_0 k^2} \operatorname{Re} H_\mathbf{k}|.$ Comparing to Figs.~\ref{fig:spectra1e2}--\ref{fig:spectra1e5}, we see that the residual is relatively small compared to typical values of the kinetic energy spectra, confirming the approximate correctness of the identity. We have excluded the slice $k_x$ where the identity breaks down. From the plots, a few locations in $\mathbf{k}$-space stand out as having relatively large residual. These appear to coincide with locations with large $\operatorname{Im} H_\mathbf{k}$ --- consistent with the fact that the same argument which led to the cross-spectral identity led us to conclude that  $\operatorname{Im} H_\mathbf{k} \simeq 0.$ These $\mathbf{k}$ are where the identity is starting to break down are those where nonlinear effects are especially important, and we expect that if the turbulence intensity were to increase, these locations would be larger and more intense. Finally, we note that while the cross-spectral identity is apparently surprisingly good at $\beta=10^5$, its utility is greatly reduced since so much of the kinetic energy is concentrated at the largest parallel scales.

\section{Conclusions and discussion}\label{sec:conc}
We have studied a simple, popular model of the solar tachocline --- $\beta$-plane MHD turbulence --- with a strong mean field in the plane, and examined in particular the critical question of momentum transport. Our simulations show three distinct regimes: an Alfv\'enic regime at small $\beta$, a Rossby regime at large $\beta$ with robust zonal flows, and a transitional regime near $\beta\simeq b_0 k_f^2$. The transitional regime is characterized by a substantial global cross-helicity and a breakdown of equipartition between the (turbulent) kinetic and magnetic energies, and thus the breakdown of Alfv\'enization. As observed in \cite{tobias2007}, by Alfv\'enizing the turbulence, the magnetic field can prevent the turbulence from acting as a negative viscosity, thus avoiding the issue raised by Gough and McIntyre. However, it is import to note that whether this is the case depends sensitively on the magnitude of $\beta$ and in particular the comparison between $\ell_{\rm MR}$ and typical scales of the turbulence.

We quantified these results with careful analytical calculations. We showed shown that, in this system, the total turbulent cross-helicity builds up to a predictable level, using a simple but robust analytic argument based on Zel'dovich's theorem. Moreover, we derived an equality relating the time-averaged cross-helicity spectrum and the time-averaged Maxwell-Reynolds stress differential, as a fairly generic consequence of separation between linear and nonlinear timescales (this cross-spectral identity should have an analog in any weakly turbulent system with multiple interacting modes). As a result, for weak turbulence, momentum transport is set by the cross-helicity, and the time-averaged turbulent resistivity vanishes. Coupling this cross-spectral identity with a careful weak turbulence closure calculation, we computed the $\beta$ scaling of the momentum transport (as measured by the real part of the Els\"asser alignment spectrum) and related quantities. This calculation quantifies the transition away from Alfv\'enic turbulence as the differential rotation intensifies.

Throughout our calculations, we assumed kinetic forcing at a small scale. This was a choice made primarily for analytic facility rather than realism, but most of our basic conclusions should be generalizable to other choices of the forcing. Similarly, a complete calculation should relax our choice that the magnetic field be aligned perpendicular to the planetary vorticity gradient.

Working in the weak turbulence regime mercifully provided analytical facility, but from a phenomenological point of view, weak turbulence was not a completely absurd assumption to impose. The tachocline supports a fairly strong toroidal field of the order $10^4$--$10^5$ G \cite{charbonneau1997}, and moreover may operate on the cusp of marginal stability \cite{garaud2001}.

The present work suggests a number of avenues for future research. A complete calculation should include the dynamics of the zonal modes, which is simply stationary in this weak turbulence picture. A strong turbulence calculation is also conspicuously absent from the present work. In the strong turbulence regime, the cross-spectral identity will break down; indeed, the very notion of linear Rossby-Alfv\'en modes ceases to be meaningful in strong turbulence. However, we still anticipate that cross-helicity could play an important role: the stationary cross-helicity is proportional to $\langle \tilde b^2 \rangle/b_0,$ which may be very large in a strongly turbulent regime. In particular, there may be an effect on the flux of magnetic potential, which is vanishing in weak turbulence, but has magnetic Reynolds number dependence \cite{cattaneo91,gruzinov94,gruzinov96} and intermittent spatial structure \cite{fan2019} when the mean field is weak. Future research should investigate the strong turbulence regime, and determine how the turbulent resistivity depends on $\beta$. Finally, it would be useful to repeat the analysis in a three-dimensional or quasi-two-dimensional setting such as SMHD and thus include effects of the inhomogeneity along the radial direction. Such a calculation should give an estimate of the turbulent emf and thus may provide some insights on the dynamo. 
\appendix
\section{Derivation of weak turbulence spectral equations} \label{app:wwt}
The derivation is based on the single-mode case presented in \cite{diamond_book}. Begin by defining the mode amplitudes $\hat \phi_\mathbf{k}^\alpha = e^{i \omega_\mathbf{k}^\alpha t} \phi_\mathbf{k}^\alpha$. Equation (\ref{eq:fieldeqs}) becomes
\be
\partial_t \hat \phi_\mathbf{k}^\alpha = \frac{1}{2} \sum_{\beta \gamma} \sum_{\mathbf{k}' + \mathbf{k}''=\mathbf{k} }  e^{i(\omega^\alpha_\mathbf{k} - \omega^\beta_{\mathbf{k}'}-\omega^\gamma_{\mathbf{k}''})t} M^{\alpha \beta \gamma}_{\mathbf{k}, \mathbf{k}' ,\mathbf{k}''} \hat \phi^{\beta}_{\mathbf{k}'} \hat \phi^{\gamma}_{\mathbf{k}''}.
\ee

We proceed via time-dependent perturbation theory, letting
\be \hat \phi_\mathbf{k}^\alpha(t) = \hat \phi_\mathbf{k}^\alpha(0) + \delta \hat \phi_\mathbf{k}^{\alpha,(1)}(t) + \delta \hat \phi_\mathbf{k}^{\alpha,(2)}(t) + \dots. \ee
(To be clear, $\delta \hat \phi_\mathbf{k}^{\alpha,(n)}(t)$ is the $n$-th order correction to $\hat \phi_\mathbf{k}^\alpha(0).$)
Then
\be
\delta \hat \phi_\mathbf{k}^{\alpha,(1)}(t) = \frac{1}{2} \sum_{\mathbf{k}' + \mathbf{k}''=\mathbf{k} } \sum_{\beta \gamma}  M^{\alpha \beta \gamma}_{\mathbf{k} ,\mathbf{k}',\mathbf{k}''} \hat \phi_{\mathbf{k}'}^{\beta}(0) \hat \phi_{\mathbf{k}''}^{\gamma}(0) \int_0^t dt' \, e^{i(\omega^\alpha_\mathbf{k} - \omega^\beta_{\mathbf{k}'}-\omega^\gamma_{\mathbf{k}''})t'}
\ee
and
\begin{align}
\delta \hat \phi_\mathbf{k}^{\alpha,(2)}(t) &= \frac12 \sum_{\beta \gamma \beta' \gamma'} \sum_{\substack{\mathbf{k}' + \mathbf{k}'' =\mathbf{k} \\ \mathbf{q}' + \mathbf{q}''=\mathbf{k}' } }   M^{\alpha \beta \gamma}_{\mathbf{k}, \mathbf{k}', \mathbf{k}''} M^{\beta \beta' \gamma'}_{\mathbf{k}' ,\mathbf{q}',\mathbf{q}''}  \hat \phi^\gamma_{\mathbf{k}''}(0) \hat \phi^{\beta'}_{\mathbf{q}'}(0) \hat \phi^{\gamma'}_{\mathbf{q}''}(0) \nonumber \\ & \quad\times \int_0^t dt' e^{i(\omega^\alpha_\mathbf{k} - \omega^\beta_{\mathbf{k}'}-\omega^\gamma_{\mathbf{k}''})t'}  \int_0^{t'} dt'' \, e^{i(\omega^\beta_\mathbf{k'} - \omega^{\beta'}_{\mathbf{q}'}-\omega^{\gamma'}_{\mathbf{q}''})t''},
\end{align}
where we have combined terms by exchanging species indices and using the symmetry of the coupling coefficients.

We are interested in the evolution of $C^{\alpha \alpha'}_\mathbf{k}(t) \equiv \langle \hat \phi^{\alpha}_\mathbf{k} \hat \phi^{\alpha' *}_\mathbf{k} \rangle$. Working to second order,
\be
\Delta C^{\alpha \alpha'}_\mathbf{k} = C^{\alpha \alpha'}_\mathbf{k}(t) - C^{\alpha \alpha'}_\mathbf{k}(0) = \langle \delta \hat \phi^{\alpha,(1)}_\mathbf{k} \delta \hat \phi^{\alpha',(1)*}_\mathbf{k} \rangle + \langle \delta \hat \phi^{\alpha,(2)}_\mathbf{k} \hat \phi^{\alpha'*}_\mathbf{k}(0) + \hat \phi^{\alpha}_\mathbf{k}(0) \delta \hat \phi^{\alpha',(2)*}_\mathbf{k} \rangle + \dots,
\ee
where we have anticipated that the first-order terms make no contribution.

Now,
\begin{align}
\Delta C^{\alpha \alpha',(1)}_\mathbf{k} &= \langle \delta \hat \phi^{\alpha,(1)}_\mathbf{k} \delta \hat \phi^{\alpha',(1)*}_\mathbf{k} \rangle \nonumber \\ &= \frac14 \sum_{\beta \gamma \beta' \gamma'}  \sum_{\substack{\mathbf{k}'+ \mathbf{k}'' = \mathbf{k} \\ \mathbf{q}'+ \mathbf{q}'' = \mathbf{k}}} M^{\alpha \beta \gamma}_{\mathbf{k}, \mathbf{k}', \mathbf{k}''} M^{\alpha' \beta' \gamma' *}_{\mathbf{k} ,\mathbf{q}' ,\mathbf{q}''} \langle \hat \phi^\beta_\mathbf{k'}(0) \hat \phi^\gamma_\mathbf{k''}(0) \hat \phi^{\beta' *}_\mathbf{q'}(0) \hat \phi^{\gamma' *}_\mathbf{q''}(0) \rangle \nonumber \\ & \quad\times \int_0^t dt' \int_0^t dt'' \, e^{i (\omega^\alpha_\mathbf{k} - \omega^\beta_\mathbf{k'} - \omega^\gamma_\mathbf{k''}) t'} e^{-i (\omega^{\alpha'}_\mathbf{k} - \omega^{\beta'}_\mathbf{q'} - \omega^{\gamma'}_\mathbf{q''}) t''}.
\end{align}

Next, we make the random phase approximation, which means we can apply Wick's theorem to the four-mode functions, and assume spatial homogeneity, giving
\begin{align}
\langle \hat \phi^\beta_\mathbf{k'}(0) \hat \phi^\gamma_\mathbf{k''}(0) \hat \phi^{\beta' *}_\mathbf{q'}(0) \hat \phi^{\gamma' *}_\mathbf{q''}(0) \rangle &= \langle \hat \phi^\beta_\mathbf{k'}(0) \hat \phi^{\beta' *}_\mathbf{k'}(0) \rangle \langle  \hat \phi^\gamma_\mathbf{k''}(0) \hat \phi^{\gamma' *}_\mathbf{k''}(0)\rangle  \delta_{\mathbf{k'} \mathbf{q'}}  \delta_{\mathbf{k''} \mathbf{q''}} \nonumber \\ & \quad + \langle \hat \phi^\beta_\mathbf{k'}(0) \hat \phi^{\gamma' *}_\mathbf{k'}(0) \rangle \langle  \hat \phi^\gamma_\mathbf{k''}(0) \hat \phi^{\beta' *}_\mathbf{k''}(0) \rangle \delta_{\mathbf{k'} \mathbf{q''}} \delta_{\mathbf{k''} \mathbf{q'}}
\end{align}
and allowing us to make several simplifications.

We need to evaluate the integral
\be
I = \int_0^t dt' \int_0^t dt'' \, e^{i \Delta \omega t'} e^{-i \Delta \omega' t''} = \frac{4 e^{i (\Delta \omega- \Delta \omega')t/2} \sin (\Delta \omega t/2) \sin (\Delta \omega't/2)}{\Delta \omega\Delta \omega'}
\ee
where $\Delta \omega= \omega^\alpha_\mathbf{k} - \omega^\beta_\mathbf{k'} - \omega^\gamma_\mathbf{k''}$ and $\Delta \omega'= \omega^{\alpha'}_\mathbf{k} - \omega^{\beta'}_\mathbf{k'} - \omega^{\gamma'}_\mathbf{k''}.$ 

We seek the limit of this integral, in a distributional sense, as $t\to \infty$. We only keep terms linear in $t$; physically, we are interested in times $\omega^{-1} < t < \gamma_{NL}^{-1}$.  The long-time limit vanishes unless $\Delta \omega=\Delta \omega'=0$, but its value depends on whether we take the $\Delta \omega' \to \Delta \omega$ limit or $t \to \infty$ limit first. The sensible choice is the former, and we obtain
\be I(t \to \infty) \simeq 2 \pi t  \delta(\Delta \omega) \delta_{\Delta \omega, \Delta \omega'}. \ee
The support of the Kronecker delta has measure zero in $(\mathbf{k'},\mathbf{k''})$ space unless $\Delta \omega = \Delta \omega'$ \emph{identically}, i.e.\ $\alpha= \alpha'$, $\beta=\beta'$, $\gamma=\gamma'$. Using these simplifications, we obtain

\be \Delta C^{\alpha \alpha',(1)}_\mathbf{k} =  \pi t \sum_{\mathbf{k}'+ \mathbf{k}'' = \mathbf{k}} \sum_{\beta \gamma} |M^{\alpha \beta \gamma}_{\mathbf{k}, \mathbf{k}', \mathbf{k}''} |^2 C^{\beta \beta}_\mathbf{k'} C^{\gamma \gamma}_\mathbf{k''} \delta(\omega^\alpha_\mathbf{k} - \omega^\beta_\mathbf{k'} - \omega^\gamma_\mathbf{k''})\delta_{\alpha \alpha'}, \ee
and, applying similar reasoning to the other terms of $\Delta C^{\alpha \alpha'}_\mathbf{k}$, we compute a second integral
\be
\lim_{t \to \infty}\int_0^t dt' e^{i \Delta \omega t'}  \int_0^{t'} dt'' \, e^{-i \Delta \omega^\prime t''} \simeq \left[i t {\cal P} \frac{1}{\Delta \omega} + \pi t \delta(\Delta \omega)\right] \delta_{\Delta \omega, \Delta \omega'}
\ee
and find
\begin{align}
\Delta C^{\alpha \alpha',(2)}_\mathbf{k} &= \langle \delta \hat \phi^{\alpha,(2)}_\mathbf{k} \hat \phi^{\alpha'*}_\mathbf{k}(0) + \hat \phi^{\alpha}_\mathbf{k}(0) \delta \hat \phi^{\alpha',(2)*}_\mathbf{k} \rangle \nonumber \\  &= t \sum_{\mathbf{k}'+ \mathbf{k}'' = \mathbf{k}} \sum_{\beta \gamma}  M^{\alpha \beta \gamma}_{\mathbf{k}, \mathbf{k}', \mathbf{k}''} M^{\beta \alpha \gamma}_{\mathbf{k}', \mathbf{k}, -\mathbf{k}''} C^{\alpha \alpha'}_\mathbf{k} C^{\gamma \gamma}_\mathbf{k''} \left( \pi \delta(\omega^\alpha_\mathbf{k} - \omega^\beta_\mathbf{k'} - \omega^\gamma_\mathbf{k''}) + i \mathcal {P} \frac{1}{\omega^{\alpha}_\mathbf{k} - \omega^\beta_\mathbf{k'} - \omega^\gamma_\mathbf{k''}}\right) \nonumber \\ & \quad + \mathrm{c.c.}', 
\end{align}
where $\mathrm{c.c.}'$ means the complex conjugate with $\alpha \leftrightarrow \alpha'$.

Finally, we obtain the claimed result (Eq.~\ref{eq:coll}). by combining $\Delta C^{\alpha \alpha',(1)}_\mathbf{k}$ with $\Delta C^{\alpha \alpha',(2)}_\mathbf{k}$ and approximating
\[
\partial_t C^{\alpha \alpha'}_\mathbf{k} \simeq \frac{\Delta C^{\alpha \alpha'}_\mathbf{k}}{t}.
\]

\section{Implementation of the stochastic forcing}\label{app:forcing}
The vorticity forcing function used in our simulations was very similar to the implementation in \cite{srinivasan}, with a couple key differences. Explicitly, we took the forcing to be a sum of Fourier modes
\be
f(\mathbf{x},t) = c \sum_{\mathbf{k}\in \mathcal{A}} e^{i (\mathbf{k} \cdot \mathbf{x} + \alpha_\mathbf{k}(t))} + \mathrm{c.c.}\ee
Here $\mathcal{A}$ is the annulus in wavevector space centered at $|k|=k_f$ with a chosen width $w=8$. The $\alpha_\mathbf{k}(t)$ is a set of phases which evolve stochastically, specifically as random walks with
\be
\alpha_\mathbf{k} (t+\delta t) = \alpha_\mathbf{k} (t) + \sqrt{\delta t} \eta_\mathbf{k},
\ee
where $\delta t$ is the integration timestep and $ \eta_\mathbf{k} \sim \mathcal{N}(0,\sigma^2)$. Averaging over ensembles and the spatial grid, and noting that $\alpha_\mathbf{k}(t)-\alpha_\mathbf{k}(t') \sim {\cal N} (0, |t-t'| \sigma^2),$ one can compute
\be
\langle f(\mathbf{x},t) f(\mathbf{x},t') \rangle = 2N c^2 e^{-|t-t'|/\tau_c},
\ee  
where $N$ is the number of wavevectors in ${\cal A}$, and $\tau_c = 2/\sigma^2$. Choosing 
\be
c= k_f \sqrt{\frac{\varepsilon}{N \tau_c}}
\ee
fixes the energy injection rate $\varepsilon=-\langle\psi f \rangle$ for sufficiently small $\tau_c$. Thus our forcing has a tuneable correlation time and energy injection rate and smoothly becomes temporally white in the limit $\tau_c \to 0$. 

In the thin annulus limit $w/k_f\to 0$ one can also compute the correlation function
\be
\langle f(\mathbf{x},t)  f(\mathbf{x}',t') \rangle = \frac{2\varepsilon k_f^2}{\tau_c} J_0(k_f |\mathbf{x}-\mathbf{x}'|)e^{-|t-t'|/\tau_c},
\ee
where $J_0$ is the Bessel function of the first kind. This leads to the $\mathbf{k}$-space representation

\be
\langle f(\mathbf{k},t) f(-\mathbf{k},t') \rangle = \frac{\varepsilon k_f }{\pi \tau_c} \delta(k-k_f) e^{-|t-t'|/\tau_c}. \ee

\section{Contribution of forcing to spectra}\label{app:forcing_wwt}
The calculation of App.~\ref{app:wwt} omitted the contribution of the forcing terms. Now, add a forcing term $f^\alpha_\mathbf{k}$ to the RHS of the equation for each $\phi^\alpha_\mathbf{k}$. Assuming separation between the linear timescale and the forcing correlation time, this will simply introduce a (symmetrized) correlator with the forcing to the equation for the two-point function:
\begin{equation}
\partial_t \langle \phi^\alpha_\mathbf{k} \phi^{\alpha'}_\mathbf{-\mathbf{k}} \rangle = \dots + \langle f^\alpha_\mathbf{k} \phi^{\alpha'}_\mathbf{-k} \rangle + {\rm c.c.}'. \end{equation}

We need to compute the correlator $ \langle f^\alpha_\mathbf{k} \phi^{\alpha'}_\mathbf{-k} \rangle.$ Neglecting nonlinear terms, we have
\[
\phi^{\alpha}_\mathbf{k} \simeq \int_0^t dt' \, f^\alpha_\mathbf{k}(t') \exp(-i \omega^\alpha_\mathbf{k} (t-t')),
\]
so
\[
\langle \phi^{\alpha}_\mathbf{k} f^{\alpha'}_\mathbf{-k} \rangle \simeq \int_0^t dt' \,\langle f^\alpha_\mathbf{k}(t') f^{\alpha'}_\mathbf{-k}(t) \rangle \exp(-i \omega^\alpha_\mathbf{k} (t-t')) = F_\mathbf{k}^{\alpha \alpha'}, 
\]
where in the last step we have assumed temporally white forcing $\langle f^\alpha_\mathbf{k}(t) f^{\alpha'}_\mathbf{-k}(t') \rangle = F^{\alpha \alpha'}_\mathbf{k} \delta(t-t')$.

For the system under study, $f^\pm_\mathbf{k} = \omega^\pm_\mathbf{k} f_\mathbf{k}/\Omega_\mathbf{k} k_f^2$, where $f_\mathbf{k}$ is the vorticity forcing. In particular, using $A= (\omega^+ \phi^- - \omega^- \phi^+)/\omega_A$ we recover in the $\tau_c \to 0$ limit
\begin{align}
\langle \tilde A_\mathbf{k} \tilde f_\mathbf{-k} \rangle &= 0.
\end{align}
Thus to good approximation we may neglect the correlator $\langle \tilde A  \tilde f \rangle.$

However, while the correlator is zero, the rms correlation is nonzero and can be estimated (for small $\beta$) as
 \[ \langle \tilde A^2 \tilde f^2 \rangle^{1/2} \sim   \frac{k_f \ell_b \varepsilon^{1/2} \langle \tilde b^2 \rangle^{1/2}}{\tau_c^{1/2}} \]
 using $\tilde f^2 \sim \varepsilon k_f^2/\tau_c$. Assuming $\langle \tilde A^2 \rangle$ equilibrates faster than the cross-helicity, the dynamics of the cross-helicity can then be modeled as an Ornstein-Uhlenbeck process 
 \[d H_t = \theta (\mu - H_t) dt + \sigma dW_t \]
 with $\theta \sim (\nu+\eta)/\ell_v \ell_b$, $\mu \sim \ell_v \ell_b \beta \langle \tilde b^2\rangle / b_0 (1+\mathrm{Pm}),$ and $\sigma \sim k_f \ell_b \varepsilon^{1/2} \langle \tilde b^2 \rangle^{1/2}.$ The asymptotic variance of this process is
 \[ \langle H_t^2 \rangle \simeq \frac{\sigma^2}{2\theta} \sim \frac{k_f^2 \ell_v \ell_b^3 \varepsilon \langle \tilde b^2 \rangle}{(\eta + \nu)}, \]
 so if
\[ \varepsilon \gg \frac{\ell_v}{\ell_b} \frac{k_{\rm MR}^4}{k_f^2}  \frac{\eta}{1+{\rm Pm}} \langle \tilde b^2 \rangle, \]
then the fluctuations in $H$ dominate its mean. For small $\beta$ we can estimate $\ell_v \sim \ell_b$ and $\langle \tilde b^2\rangle\sim \varepsilon/(\eta+\nu)k_f^2, $ whence the condition for the mean cross-helicity to not be dominated by fluctuations is
\[
k_{\rm MR}^2 \gtrsim k_f^2 (1+{\rm Pm}).
\]

\acknowledgements{The authors thank S.\ M.\ Tobias, D.\ W.\ Hughes, and C.-C.\ Chen for useful discussions. We also acknowledge stimulating interactions with participants of the 2019 Festival de Th\'eorie and the 2021 KITP program Staircase 21. KITP is supported in part by the National Science Foundation under Grant No.\ NSF PHY-1748958. This work used the Comet cluster at the San Diego Supercomputing Center (SDSC) (through allocation TG-PHY190014), which is part of the Extreme Science and Engineering Discovery Environment (XSEDE) \cite{xsede}. XSEDE is supported by National Science Foundation grant number ACI-1548562. We also gratefully acknowledge the computing resources provided on Bebop and Blues, high-performance computing clusters operated by the Laboratory Computing Resource Center at Argonne National Laboratory. We thank Olle Heinonen for providing access to the LCRC resources. The research was supported by the U.S. Department of Energy, Office of Science, Office of Fusion Energy Sciences under Award Number DE-FG02-04ER54738. It also received funding from the European Union's H2020 Program under grant agreement No.\ 882340.}

%


\end{document}